\documentclass[universe,aricle,accept,pdftex,moreauthors]{Definitions/mdpi} 

\usepackage[labelformat=simple]{subcaption}

\DeclareCaptionLabelFormat{subcaptionlabel}{\normalfont(\textbf{#2}\normalfont)}
\firstpage{1} 
\pubvolume{10}
\issuenum{10}
\articlenumber{162}
\pubyear{2024}
\copyrightyear{2024}
\externaleditor{Academic Editor: Pier Stefano Corasaniti}
\datereceived{23 February 2024}
\daterevised{21 March 2024} 
\dateaccepted{27 March 2024} 
\datepublished{29 March 2024} 
\hreflink{https://doi.org/\linebreak 10.3390/universe10040162
} 
\pdfoutput=1

\DeclareFontEncoding{LS1}{}{}
\DeclareFontSubstitution{LS1}{stix}{m}{n}
\DeclareSymbolFont{scr}{LS1}{stixscr}{m}{n}
\DeclareSymbolFontAlphabet{\mathscr}{scr}

\newcommand{\PC}[1]{\ensuremath{\left(#1\right)}}
\newcommand{\PR}[1]{\ensuremath{\left[#1\right]}}
\Title{Dynamical Analysis of the Redshift Drift in FLRW Universes}

\TitleCitation{Dynamical Analysis of the Redshift Drift in FLRW Universes}

\Author{{Francisco} 
 S. N. Lobo $^{1,2}$\orcidA{}, {José Pedro Mimoso} 
 $^{1,2}$\orcidB{}, Jessica Santiago $^{3,}$*\orcidC{}, and Matt Visser $^{4}$\orcidD{}}

\AuthorNames{Francisco S. N. Lobo, José Pedro Mimoso, Jessica Santiago and Matt Visser}

\AuthorCitation{{Lobo,} 
 F.S.N.; Mimoso, J.P.; Santiago, J.; Visser, M.}

\address{%
$^{1}$ \quad Instituto de  Astrof\'{i}sica e Ci\^encias do Espa\c co, 
{Faculdade de Ci\^{e}ncias, Universidade de Lisboa,} 
 Campo Grande, Edif\'{\i}cio C8, 1749-016 Lisboa, Portugal; fslobo@fc.ul.pt (F.S.N.L.);~jpmimoso@fc.ul.pt (J.P.M.)\\
$^{2}$ \quad Departamento de F\'{i}sica, {Faculdade de Ci\^{e}ncias, Universidade de Lisboa}, Campo Grande, Edif\'{\i}cio C8, 1749-016 Lisboa, Portugal\\
$^{3}$ \quad Leung Center for Cosmology and Particle Astrophysics, National Taiwan University, Taipei 10617, Taiwan\\
$^{4}$ \quad School of Mathematics and Statistics, Victoria University of Wellington, P.O. Box 600, \linebreak  Wellington 6140, New Zealand;~matt.visser@sms.vuw.ac.nz}

\corres{Correspondence: jessicasantiago@ntu.edu.tw
}

\abstract{Redshift drift is the phenomenon whereby the observed redshift between an emitter and observer comoving with the Hubble flow in an expanding FLRW universe will slowly evolve---on a timescale comparable to the Hubble time. 
In a previous {article},
three of the current authors performed a cosmographic analysis of the redshift drift in an FLRW universe, temporarily putting aside the issue of dynamics (the Friedmann equations). 
In the current article, we add dynamics while still remaining within the framework of an exact FLRW universe. 
We developed a suitable generic matter model and 
applied it to both standard FLRW and various dark energy models. Furthermore, we present an analysis of the utility of alternative cosmographic variables to describe the redshift drift data.}

 
\begin{document}

\section{Introduction}
\label{S:introduction}

The concept of ``redshift drift'' (RD) dates back (at least) some 60 years, to 1962, arising in coupled papers by  Sandage~\cite{Sandage:1962} and McVittie~\cite{McVittie:1962}. 
Relatively little direct follow-up work took place in the 20th century, with Loeb's 1998 article~\cite{Loeb:1998} as a stand-out exception. However, with technological advances and new observational surveys on the horizon, the possibilities of measuring RDs have become much more concrete \cite{Liske:2008,Linder:2008,Quercellini:2010, Steinmetz:2008,Kim:2014,Killedar:2009, Marcori:2018,Alves:2019, Lu:2022,Dong:2022,Covone:2022, Chakrabarti:2022, Lu:2021, Martins:2021, Lazkoz:2017}. 
The basic idea is this: If in any FLRW universe emitter and observer are comoving with the Hubble flow, then the null curve connecting them slowly evolves on a timescale set by the Hubble parameter; this implies that the redshift is slowly evolving. In any FLRW universe, the key result is
\cite{Sandage:1962,McVittie:1962,Loeb:1998,Linder:2008}:
\begin{equation}
\label{keyeq}
    \dot z = (1+z) H_0 - H(z).
\end{equation}

Measuring this effect will certainly be a challenging enterprise, with typical estimates suggesting the need for a decade-long observational window. 
Starting from an estimated detection time of a couple of decades---using the first observational feasibility study of the Extremely Large Telescope (ELT) \cite{Liske:2008}---recent experimental proposals suggest a detection time as low as 6 years~\cite{Eikenberry:2019} (though the constraints provided on cosmological parameters could potentially be greatly diminished by the time-reduction~\cite{Esteves:2021}). Furthermore, other future prospects from RD measurements
are proposed to test the cosmological principle \mbox{(i.e., isotropy} and homogeneity) by taking into account large-scale structures and distinguishing between non-FLRW cosmological models \cite{Uzan:2008,Koksbang:2022,Koksbang:2015,Quartin:2009xr,Yoo:2010hi,Geng:2015ara,Calabrese:2013lga,Mishra:2012vi,Li:2021}.
More boldly, some authors have recently speculated on what might be do-able with millennia-long observational windows~\cite{Loeb:2022}.

In~\cite{Lobo:2020}, three of the current authors performed a general cosmographic analysis 
(for additional background, see~\cite{Dunsby:2015, Visser:2003, Gibbons:2008, Visser:2004,Aviles:2012, Cattoen:2007b,Cattoen:2008,  jerk3, Busti:2015,  Lusso:2020, Hu:2022, Yang:2020, Capozziello:2020}),
both in terms of the regular $z$-redshift and in terms of the $y$-redshift, defined by
\begin{equation}
1-y = {a\over a_0} = {1\over 1+z},
\end{equation}
so that for $z\in[0,\infty)$, one has $y\in[0,1)$. 
There we have proved the closely related exact result:
\begin{equation}
\dot y = (1-y)\left\{ H_0 - (1-y) H(y)\right\}.
\end{equation}
{We}  shall now analyze and extend these and related results by using the dynamical Friedmann equations of general relativistic cosmology.

Without choosing the exact composition of the universe, one can rewrite the RD in terms of a suitably defined density parameter, $\Omega(z)$ (one that includes the effect of spatial curvature), as
\begin{equation}
\dot z  = \left\{ (1+z) -\sqrt{\Omega(z)}\right\}H_0;  \qquad \qquad  (\Omega_0\equiv1). 
\end{equation}
{Similarly}, for the $y$ redshift, we have
\begin{equation}
\dot y  = (1-y) \left\{ 1 -(1-y)\sqrt{\Omega(y)}\right\}H_0;  \qquad \qquad  (\Omega_0\equiv1). 
\end{equation}

From this, we can proceed by building a physically plausible matter model for $\Omega(z)$ (or equivalently $\Omega(y)$) and investigate its properties. We will allow for an arbitrary admixture of non-interacting components, all individually satisfying linear equations of state \mbox{$p_i = w_i \, \rho_i$.} Such a model is general enough to include $\Lambda$-CDM and many variants thereof but is simple enough to allow explicit calculations of $\Omega(z)$ or, equivalently, $\Omega(y)$ and its various approximations. 

We also present an RD analysis applied to different dark energy (DE) models, such as $w_0$CDM, the linear model, BAZS equation of state, CPL, logarithmic evolution, and a couple of interactive models. This is followed by a discussion on whether RD data have the power to distinguish distinct equations of state for dark energy or not.

The article is outlined as follows: In Section~\ref{S:z-redshift}, we develop the 
notation, set the stage, and subsequently develop a dynamical analysis in terms of the usual $z$-redshift.
We also present the relations between the RD signal peak, $z_{peak}$, and $z_{equality}$, and the turning point of the acceleration rate of the universe, $q=0$, for the $\Lambda$CDM case.
In Section~\ref{S:dark_energy_models}, we start by introducing the dark energy models discussed in this work, followed by the predicted RD signal of each model for different values of the relevant free parameters. We then proceed in Section~\ref{S:DE_discussion} to discuss the power of redshift data in distinguishing different DE models from each other.
In Section~\ref{S:y-redshift}, we move towards a cosmographic analysis, presenting the general results in terms of the $y$-redshift defined by $y = {z/(1+z)}$. 
Other auxiliary variables for describing the redshift are then presented in a table in Section~\ref{S:other_auxiliary_variables}.
Finally, we conclude in Section \ref{S:Conclusions}.

\section{Dynamics of the RD in Terms of \boldmath{$z$}}
\label{S:z-redshift}

As is well known, the dynamical behavior of the Friedmann (FLRW) cosmological models is determined by two {equations:} 
\begin{itemize}
\item[(i)] The second-order Raychaudhuri equation (or second Friedmann equation):
\begin{equation}
\frac{\ddot a}{a} =-\frac{\kappa^2 (\rho+3p)}{6} \label{eq-Raychaudhuri}\;,
\end{equation}
where $\kappa^2=8\pi G_N$ and $c=1$. Here, $\rho$ and $p$ are the energy density and the pressure of the matter content of the universe, respectively, described by an isotropic perfect fluid. Note that we have adopted the simplification of absorbing the cosmological constant, if present, into the stress-energy tensor.
\item[(ii)] The first Friedmann equation:
\begin{equation}
\left(\frac{\dot a}{a}\right)^2 =-\frac{k}{a^2}+ \frac{\kappa^2 \rho}{3}\; . 
\label{eq-Friedmann}
\end{equation}
The latter equation acts as a first integral of Equation (\ref{eq-Raychaudhuri}) and constrains the solutions in connection with the possible spatial curvature cases set by $k=0,\pm 1$. 
\end{itemize}
Furthermore, it is useful to absorb the spatial curvature term into the density by defining
\begin{equation}
\rho_\mathrm{k} = -{3k\over 8\pi a^2}; \qquad 
\rho_\mathrm{effective} = \rho + \rho_\mathrm{k}.
\end{equation}
{Similarly,} it is useful to define
\begin{equation}
p_\mathrm{k} = {k\over 8\pi a^2}; \qquad 
p_\mathrm{effective} = p + p_\mathrm{k}.
\end{equation}
{In} this way, we have
\begin{equation}
H^2 = \left(\frac{\dot a}{a}\right)^2 = \frac{\kappa^2 \rho_\mathrm{effective}}{3}
\qquad\text{and}\qquad
\frac{\ddot a}{a} =-\frac{\kappa^2 (\rho_\mathrm{effective}+3p_\mathrm{effective})}{6}\;, \label{E: FE2}
\end{equation}
where the Raychaudhuri equation can be rewritten in terms of $H$ and $\omega_\mathrm{effective}=  p_\mathrm{effective}/$\linebreak  $\rho_\mathrm{effective}$ as
\begin{eqnarray}
\dot H + H^2 &=& -  \kappa^2\left(\frac{1+3\omega_\mathrm{effective}}{6}\right)\,\rho_\mathrm{effective} \;, \label{E: FE1}
\end{eqnarray}
giving us
\begin{equation}
    \dot H = -  \kappa^2\left(\frac{1+\omega_\mathrm{effective}}{2}\right)\,\rho_\mathrm{effective} \label{E: FE3}
\end{equation}
{Now,} making use of Equation \eqref{E: FE2}, one can rewrite $H(z)$ as
\begin{equation}
H = H_0 \; \sqrt{\rho_\mathrm{effective} \over \rho_\mathrm{effective,0}},
\end{equation}
where $\rho_\mathrm{effective}$ and $\rho_\mathrm{effective,0}$ represent the energy density at redshift $z$ and at the present time, respectively. 
Introducing the appropriate notion of critical density (sometimes called Hubble density) and Omega (or density) parameter
\begin{equation}
\rho_\mathrm{crit} = {3H_0^2\over\kappa^2}
=\rho_\mathrm{effective,0}; \qquad \qquad
\Omega = {\rho_\mathrm{effective}\over \rho_\mathrm{crit}};
\end{equation}
respectively, we see that these definitions automatically imply $\Omega_0\equiv1$, and hence, $H(z)= H_0\; \sqrt{\Omega(z)}$.
Thence, for the RD, even before choosing a specific cosmological model, we have the quite general exact (FLRW) result:
\begin{equation}
\label{E: zredshiftdrift}
\dot z  = \left\{ (1+z) -\sqrt{\Omega(z)}\right\}H_0;  \qquad \qquad  (\Omega_0\equiv1). 
\end{equation}
{The} RD will exhibit a zero whenever
\begin{equation}
    \Omega(z_*)= (1+z_*)^2.
\end{equation}
{One} obvious (trivial) root occurs at $z_*=0$. 
We shall soon see that, typically, there will be at least one other nontrivial root.
Again, we emphasize that, up to this point, all quoted results are exact, at least within the context of FLRW spacetime.

\subsection{Choosing a Specific Cosmological Model}

\enlargethispage{20pt}
If we can approximate the cosmological fluid by a collection of $N$ non-interacting fluid components with individual, strictly linear equations of state (EOS) $p_i = w_i \, \rho_i$, then 
\begin{equation}
\label{E:Omega0}
\Omega(z) =  \sum_{i=1}^N \Omega_{0i} (1+z)^{3(1+w_i)};
\qquad\qquad
\sum_{i=1}^N \Omega_{0i} = 1;
\end{equation}
and so
\begin{equation}
\label{E: H(z)Omega}
H(z) = H_0 \; \sqrt{ \sum_{i=1}^N \Omega_{0i} (1+z)^{3(1+w_i)} };
\qquad\qquad
\sum_{i=1}^N \Omega_{0i} = 1.
\end{equation}
{Note} that we are explicitly allowing possible spatial curvature, so both $k$ and $\Omega_k$ are allowed to be nonzero, with the corresponding value of $w$ being  $w_k = -1/3$. Once you allow nonzero $\Omega_k$, then the sum of all the $\Omega_{0i}$ is, by definition, set to unity.
We can characterize the components of this cosmological model as follows:
\begin{equation}
    w_i \;\; \left\{ \;\;\begin{array}{cl} 
    > -1/3 & \quad\hbox{ matter-like (dust, radiation, etc.); }\\
    = -1/3 & \quad\hbox{ spatial curvature (the marginal case); }\\
    < -1/3 & \quad\hbox{ dark-energy-like (quintessence, cosmological constant, etc.). }
    \end{array} \right.
\end{equation}
{Typically}, one has $w_i\in[-1,+1]$ or even $w_i\in[-1,+1/3]$, but such a restriction is not absolutely necessary.
Note that the ``matter-like'' components ($w_i > -1/3$) dominate at early times (small $a$), while the ``dark-energy-like'' components ($w_i < -1/3$) dominate at later times (large $a$). We shall now investigate the implications of this model for $\Omega(z)$ in some detail. 
We shall find it advantageous to work with weighted moments of the $w$-parameters (for an akin definition of an averaged adiabatic index $\gamma=1+w$, see \cite{Madsen:1992}). Specifically, at the current epoch, we define
\begin{equation}
    \langle w^n \rangle_0 = 
    {\sum_{i=1}^N \Omega_{i0} \; w_i^n\over\sum_{i=1}^N \Omega_{i0} } =
    \sum_{i=1}^N \Omega_{i0} \; w_i^n,
\end{equation}
while at redshift $z$, we define
\begin{equation}
    \langle w^n \rangle_z = 
    {\sum_{i=1}^N \Omega_{i}(z) \; w_i^n
    \over\sum_{i=1}^N \Omega_{i}(z) } =
    {\sum_{i=1}^N \Omega_{i0} (1+z)^{3(1+w_i)} \; w_i^n \over 
    \sum_{i=1}^N \Omega_{i0} (1+z)^{3(1+w_i)}}.
\end{equation}
{Note} that Taylor series expansions of $\Omega(z)$ will mathematically converge only for $|z|<1$ and will be astrophysically most useful only for $0\leq z\ll1$.
Furthermore, within the context of our $\{\Omega_{0i},w_i\}$ matter model, for $\omega_\mathrm{effective}=  p_\mathrm{effective}/\rho_\mathrm{effective}$, we have
\begin{equation}
    \omega_\mathrm{effective} = {\sum_{i=1}^N \Omega_{i}(z) \; w_i
    \over\sum_{i=1}^N \Omega_{i}(z) } = \langle w \rangle_z\;.
\end{equation}
{The} dominance of some individual component, $j_\ast$, with regard to the others can be characterized by  $\Omega_{j_\ast}(z) >1/2$ (this follows trivially from $\Omega_{j_\ast} (z) > \sum_{i\neq {j_\ast}} \Omega_{i} (z)=(1- \Omega_{j_\ast}(z)) $).
Furthermore, the emergence of late-time accelerated expansion happens, of course, for $\langle w\rangle_z <-1/3$ (for some $z<z_{\rm crit}$ and, in particular, for $z\to 0$). 
We shall have more to say on these issues later on.

Given the convergence issue with the $z$-redshift expansions for $z>1$---namely all the regions of interest for RD experiments---one can resort, for example, to the $y$-redshift:
\begin{equation}
\label{E:Omega0 y}
\Omega(y) =  \sum_{i=1}^N \Omega_{0i} (1-y)^{-3(1+w_i)};
\qquad\qquad
\sum_{i=1}^N \Omega_{0i} = 1;
\end{equation}
whence
\begin{equation}
\label{E: H(z)Omega y}
H(y) = H_0 \; \sqrt{ \sum_{i=1}^N \Omega_{0i} (1-y)^{-3(1+w_i)} };
\qquad\qquad
\sum_{i=1}^N \Omega_{0i} = 1.
\end{equation}
{Again}, the Taylor series expansions of $\Omega(y)$ will mathematically converge only for $|y|<1$ and will be astrophysically most useful only for $0\leq y\ll1$. Fortunately, $0\leq y<1$ covers the entire physically relevant region $0\leq z <\infty$. Again, it is important to highlight this is the reason why so much effort is put into working with the $y$-redshift (or its variants). 

\subsubsection{Generic Model}
%
By inserting \eqref{E: H(z)Omega} into \eqref{E: zredshiftdrift} (this is still exact in FLRW with this particular model for the cosmological EOS), we obtain the generic result:
\begin{equation}
\label{E:generic_model}
\dot z = H_0 \left\{ (1+z) - \sqrt{ \sum_{i=1}^N \Omega_{0i} (1+z)^{3(1+w_i)} } \right\};
\qquad\qquad
\sum_{i=1}^N \Omega_{0i} = 1.
\end{equation}
{Observationally,} one now merely needs to fit the $\{\Omega_{0i},w_i\}$ to the empirical data. 
In contrast, a theorist need only fit the $\{\Omega_{0i},w_i\}$ to their preferred toy model.

\subsubsection{$\Lambda$-CDM}
\begin{figure}[H]

    \includegraphics[scale=0.5]{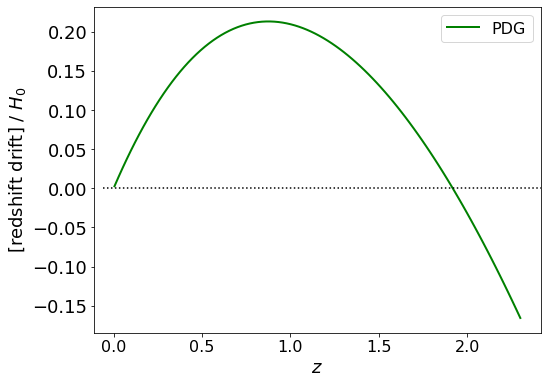}
    \caption{Redshift drift (per $H_0$) for four-component $\Lambda$-CDM, estimated from the PDG data.}
    \label{F:one}
\end{figure}

For a general four-component $\Lambda$-CDM model, when explicitly allowing the inclusion of both spatial curvature $\Omega_k$ and radiation $\Omega_r$ components, one has
\begin{equation}
H = H_0 \sqrt{ \Omega_\Lambda + \Omega_k  (1+z)^2 
+ \Omega_m (1+z)^3 + \Omega_r (1+z)^4},
\end{equation}
with
\begin{equation}
\Omega_\Lambda + \Omega_k  
+ \Omega_m  + \Omega_r =1.
\end{equation}
{This} particular model is commonly believed to be an accurate representation of the evolution of our own universe from the current epoch to at least as far back as the surface of 
last scattering (the CMB) at $z\approx 1100$. Thence, by eliminating $\Omega_k$, one has
\begin{equation}
H = H_0 \sqrt{ \Omega_\Lambda + (1-\Omega_\Lambda -\Omega_m -\Omega_r) (1+z)^2 
+ \Omega_m (1+z)^3 + \Omega_r (1+z)^4}\;,
\end{equation}
which leads to the RD equation in terms of the cosmological parameters:\vspace{-12pt}
\begin{adjustwidth}{-\extralength}{0cm}
\centering 
\begin{equation}
\label{E:zdot}
\dot z = H_0 \left\{ (1+z) - 
\sqrt{ \Omega_\Lambda + (1-\Omega_\Lambda -\Omega_m -\Omega_r) (1+z)^2 
+ \Omega_m (1+z)^3 + \Omega_r (1+z)^4}\right\}\;.
\end{equation}
\end{adjustwidth}
Using the latest PDG 2022 data, 
the density parameters at the current epoch are estimated to be~\cite{PDG:2022, Workman:2022} 
\begin{equation}
  \Omega_\Lambda= 0.685(7), \qquad \Omega_m = 0.315(7), \qquad
\Omega_r = 5.38(15)\times 10^{-5},  \qquad \Omega_k = 0.0007(19). \nonumber  
\end{equation}
By using those data to plot the RD vs. $z$, we can see (Figure \ref{F:one}) that the RD has a maximum at $z\approx 0.875$, where $\dot z \approx 0.213 H_0$, which then, subsequently, has a
zero at $z\approx 1.918$, beyond which the RD becomes negative---a well-known result due to the fact that the universe is matter-dominated for $z \gtrsim 2$. 
While the existence of this peak in the RD is tolerably well known \cite{Yoo:2010hi,Koksbang:2015},
we will have considerably more to say on this point later.

\subsubsection{$\Lambda$-CDM (Simplified Two-Component Version)}

One can also further simplify the discussion above
by making the plausible approximations of setting $\Omega_k\longrightarrow0$ and $\Omega_r\longrightarrow0$, obtaining the simplified two-component model:
\begin{equation}
H(z) = H_0 \sqrt{ \Omega_\Lambda
+ \Omega_m (1+z)^3};
\qquad
\Omega_\Lambda + \Omega_m  =1.
\end{equation}
{Thence}
\begin{eqnarray}
\dot z &=& H_0 \left\{ (1+z) - 
\sqrt{ \Omega_\Lambda 
+ \Omega_m (1+z)^3 }\right\}; \qquad 
\Omega_\Lambda + \Omega_m  =1.\nonumber\\
\label{E:LCDM}
&=& H_0 \left\{ (1+z) - 
\sqrt{ \Omega_\Lambda 
+ (1-\Omega_\Lambda) (1+z)^3 }\right\}.
\end{eqnarray}
{Setting} $\Omega_k\longrightarrow0$, that is, $k/a_0^2 \longrightarrow 0$, is a common simplifying assumption within the framework of cosmological inflation. Setting $\Omega_r\longrightarrow0$ is a reasonable approximation for the relatively recent universe (say $z<100$), as long as one does not try to extrapolate all the way back to the CMB.

Just like before, one finds that $\dot z$ has a zero and switches sign when $z=z_\mathrm{critical}$, analytically given by
\begin{equation}
z_\mathrm{critical} = {{3\over2}\Omega_\Lambda -1+ \sqrt{\Omega_\Lambda\left(1-{3\over4}\Omega_\Lambda\right)}\over 1-\Omega_\Lambda}. 
\end{equation}
{Equivalently,} in terms of $\Omega_m$, we have
\begin{equation}
z_\mathrm{critical} = {1-3\Omega_m 
+ \sqrt{(1 - \Omega_m)(1+3 \Omega_m) }\over 2\Omega_m}. 
\end{equation}
{As} one can see, the location where the RD vanishes depends only on the amount of dark energy \emph{versus} matter present in the universe for a two-component dust-based flat $\Lambda$CDM.

Similarly, one can find the position $z_\mathrm{peak}$ of the maximum signal of the redshift-drift by solving $d[\dot{z}(z)]/d z=0$. From (\ref{E:LCDM}), the condition for finding the peak is 
\begin{equation}
    1 - {3(1-\Omega_\Lambda) (1+ z_\mathrm{peak})^2 
    \over 2 \sqrt{ \Omega_\Lambda + (1-\Omega_\Lambda) (1+z_\mathrm{peak})^3 }} =0.
\end{equation}
{This} can be rearranged to yield a quartic polynomial:
\begin{equation}
    9 (1-\Omega_\Lambda)^2 (1+ z_\mathrm{peak})^4 - 4 (1-\Omega_\Lambda) (1+ z_\mathrm{peak})^3 
    - 4 \Omega_\Lambda =0.
    \label{E: zp}
\end{equation}
{While} the exact roots can certainly be found algebraically, they are too complicated to be worth writing down explicitly. However, a numerical evaluation of the physically relevant root is trivial.

\begin{figure}[H]

    \includegraphics[scale=0.5]{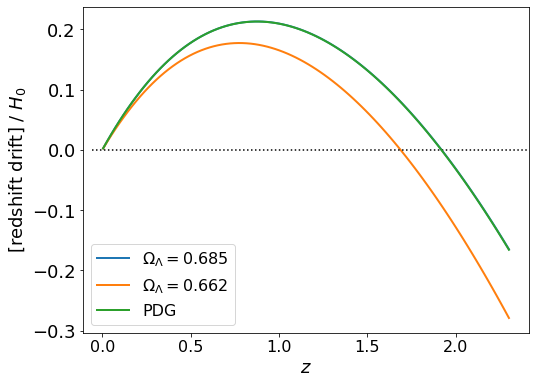}
    \caption{Redshift drift magnitude (per $H_0$) for two-component $\Lambda$-CDM for different values of $\Omega_{\Lambda}$. Here, we have taken the values for $\Omega_{m} = 0.315\pm 0.007$ given by the Planck2018 data  release~\cite{Planck2018} (blue curve) and $\Omega_m = 0.338 \pm 0.018$ from Pantheon+~\cite{Pantheon+} (orange curve) plus the values previously given by the PDG data (green curve). Note that since both Planck (assuming $\Omega_k = \Omega_r = 0$) and PDG have exactly the same value for $\Omega_{\Lambda}$, the two curves appear to exactly coincide.}
    \label{fig:RD}
\end{figure}

As an example, in Figure \ref{fig:RD}, we see that for $\Omega_\Lambda = 0.685$, one finds $z_\mathrm{critical}\approx 1.92$, with the peak at $z_\mathrm{peak}\approx 0.876$, while for $\Omega_\Lambda = 0.662$, one finds $z_\mathrm{critical}\approx 1.69$, with the peak at $z_\mathrm{peak}\approx 0.775$.
This is qualitatively compatible with what we saw happening in Figure~\ref{F:one} for the four-component model based on the PDG data. 

\vspace{-3pt}

The importance of these observations lies in the fact that, apart from being able to probe the various matter components of the universe, one should devote some care when selecting celestial objects for potential investigation. 
Note, for example, that collecting data in the region $z\in (1.7,2.0)$ is contra-indicated since the RD is approximately zero in this region. Moreover, while the region $z\in (0.5, 1.3)$ allows for a possible \emph{{local}} 
 maximum of the RD, going for redshifts higher than $z \approx 2.3$ will certainly guarantee better results in terms of the strength of the signal.
Besides the experimental difficulties in measuring such a weak signal, unlike a large portion of the cosmological data available, which depends on some sort of calibration of the distance ladder, and the RD measurements are not directly affected by this problem. 

\subsubsection{The Relation between $z_{peak}$, $z_{equality}$, and $q=0$}

One interesting exercise is to understand the relation between the position, $z_{peak}$, of the peak signal in the RD curve and the composition of the universe when this occurs. Moreover, one may consider that the position of the peak occurs at a time when the universe was still matter-dominated. In order to see this in the simplified two-component model, note that
\begin{equation}
    \Omega_\Lambda(z) = \Omega_\Lambda; \qquad \Omega_m(z) = \Omega_m\;(1+z)^3.
\end{equation}
{Thus,} matter dominates over the cosmological constant for
\begin{equation}
    z > z_{equality} 
    = \left(\Omega_\Lambda\over\Omega_m\right)^{1/3} - 1 
    = \left(\Omega_\Lambda\over1-\Omega_\Lambda\right)^{1/3} - 1,
    \label{E: zeq}
\end{equation}
which, when using the PDG data, gives us $z_{equality}\approx 0.296$. Given that $z_{peak}\approx 1.88$,
it is clear that this happened at values of $z$ higher than when the universe became cosmologically constant dominated. 
Perhaps counter-intuitively, although the universe has to be dominated by a cosmological constant \emph{now} in order for the redshift peak to \emph{exist}, the light that comes to us at the peak of the RD curve was emitted before this transition happened. 

\begin{figure}[H]
\begin{subfigure}{0.48\textwidth}
\includegraphics[width=\linewidth]{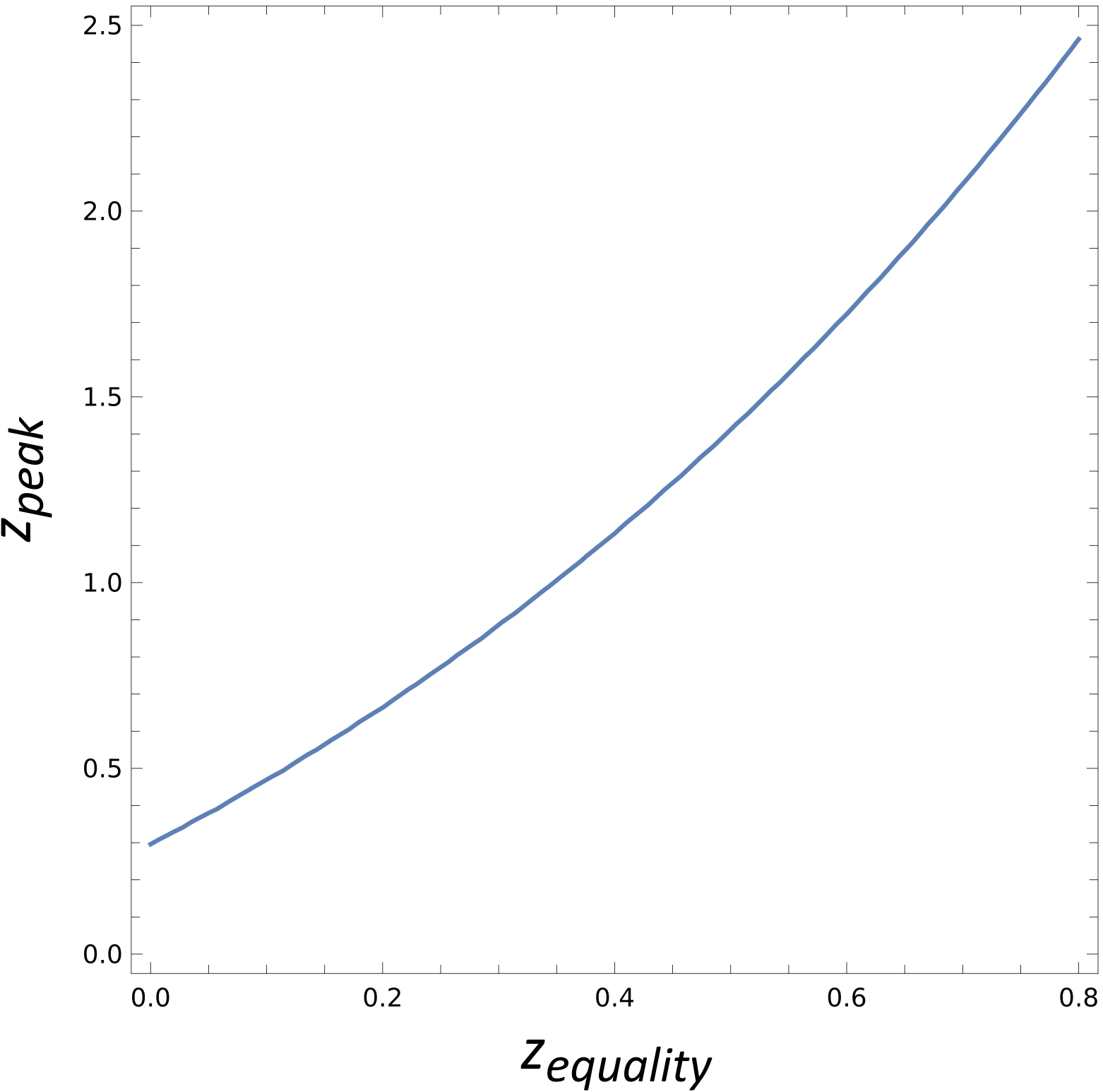}
\caption{} \label{fig:zpeakzeq}
\end{subfigure}~~
\begin{subfigure}{0.48\textwidth}
\includegraphics[width=\linewidth]{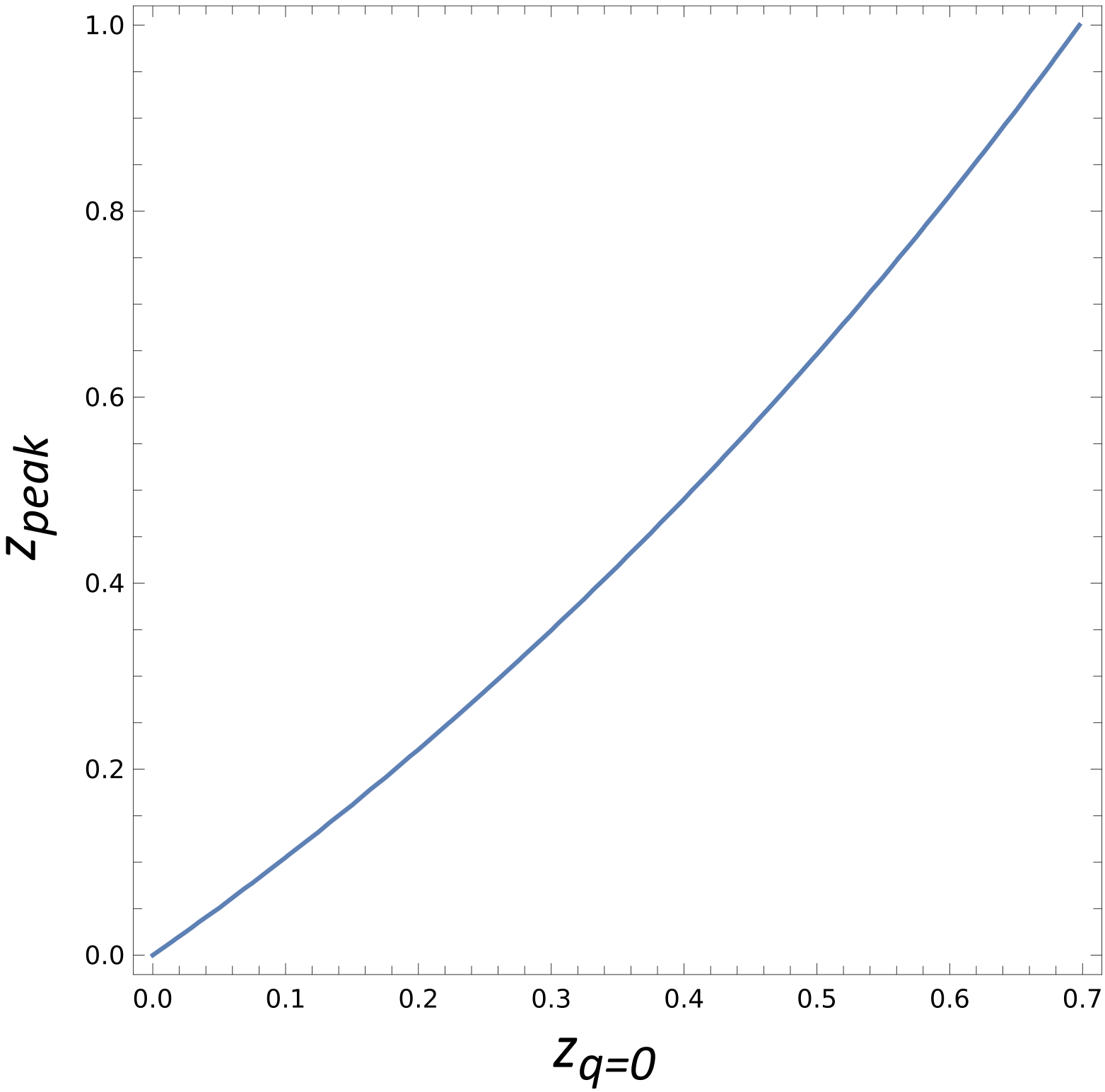}
\caption{} 
\end{subfigure}
\caption{(\textbf{a}) {Relation} 
 between $z_{peak}$ and  $z_{equality}$. Note how a peak would still be present in the RD data even if equality between $\Omega_{\Lambda}$ and $\Omega_m$ has not yet been reached. 
 (\textbf{b}) Comparison between $z_{peak}$ and $z_{q=0}$ for a two-component flat $\Lambda$CDM model.} \label{fig:zpeakzq}
\end{figure}

As a way to visualize the relation between $z_{equality}$ and $z_{peak}$, one can substitute \eqref{E: zeq} into \eqref{E: zp}, obtaining the following relation:
\begin{equation}
    4\,{{ Z_{eq}}}^{6}+ \left( 4\,{{Z_{peak}}}^{3}+4 \right) {{Z_{eq}}}^{3}
-9\,{{Z_{peak}}}^{4}+4\,{{Z_{peak}}}^{3}=0\;,
\end{equation}
where $Z_{eq} = 1+z_{equality}$ and $Z_{peak} = 1+z_{peak}$. Note, again, that this relation is only valid assuming the simplified two-component model of flat $\Lambda$CDM.
In Figure \ref{fig:zpeakzq}, is possible to see that, even if equality was only reached today, we would still see a peak in the RD at $z_{peak}\approx 0.297$.

\vspace{-3pt}

We could also try to estimate the epoch at the end of matter domination in a different way by considering the redshift at which the deceleration parameter $q(z)$ crosses zero.
From the Friedmann equations, we know that $\ddot a \propto \rho+3p$; so, the condition for the zero deceleration parameter $q(z)=0$ is $\rho+3p=0$. However, for our general matter model,
\begin{equation}
[\rho+3p](z) = \rho_0 \sum_{i=1}^N (1+3w_i) \Omega_i(z) =
\rho_0 \sum_{i=1}^N (1+3w_i) \Omega_{0,i}(1+z)^{3(1+w_i)}.
\end{equation}
{So,} for the two-component case,
\begin{equation}
[\rho+3p](z) = 
\rho_0 \left\{-2 \Omega_\Lambda +(1-  \Omega_\Lambda) (1+z)^3   \right\}.
\end{equation}
{This} is zero when
\begin{equation}
2 \Omega_\Lambda =(1-  \Omega_\Lambda) (1+z)^3 .
\end{equation}
{Thence}
\begin{eqnarray}
\label{E: zq0}
z_{q=0} &=& 
\left( 2 \Omega_\Lambda\over 1-\Omega_\Lambda\right)^{1/3} - 1\\
&=&  2^{1/3}z_{equality} - (1 -2^{1/3}) \;\;\approx \;\; 2^{1/3}z_{equality} - 0.666 \nonumber
\end{eqnarray}
{For} $\Omega_\Lambda \approx 0.685$, one has $z_{q=0} \approx 0.6323$. Furthermore, we can investigate the relation between $z_{q=0}$ and $z_{peak}$. This can be carried out by replacing \eqref{E: zq0} into \eqref{E: zp}, which gives us the following:
\begin{equation}
  4 Z_{q}^6 +(8 Z_{peak}^3 +8)Z_q^3 - 36 Z_{peak}^4 + 16 Z_{peak}^3=0
\end{equation}
where $Z_{q} = 1+z_{q=0}$ and $Z_{peak} = 1+z_{peak}$. 
As can be seen in Figure \ref{fig:zpeakzq}b, the offset existing between $z_{equality}$ and $z_{peak}$ doesn't exist between $z_{q=0}$ and $z_{peak}$. Putting it bluntly, if $q<0$, then we must undoubtedly see a peak somewhere in the RD data. 

All three calculations have a slightly different physical meaning but are still intrinsically correlated. Regardless of the method, however, the switchover from matter to (a cosmological constant form of) dark energy is certainly at $z < 1$.

\subsubsection{Clarifying a Potential Mis-Application to the CMB}

As an interesting exercise, one could \emph{{very}} naively try to apply the RD reasoning to the CMB. This would be a terrible \emph{{misapplication}}, resulting in extremely wrong results. It is important, however, 
to understand \emph{{why}} the correct result cannot be reproduced via an RD
analysis and, of course, how to do it properly.

As is well known, the CMB originated at $z\approx 1100$. If one insisted on applying (\ref{E:LCDM}) to this case, they would roughly (and again, \emph{{very}} naively) obtain something around
\begin{equation}
\left.\dot z\right|_\mathrm{CMB} \approx -18,900 \; H_0 \approx 10^{-6}/\hbox{year}.
\end{equation}
{However,} this estimate is grossly misleading. 
The basic problem is that, as presented above, this calculation would only be relevant to an emitter that was comoving with the Hubble flow at the time of last scattering.

However, the surface of 
last scattering does not comove with the Hubble flow; certainly, the surface of 
last scattering is not a physical object locally at rest in the FLRW spacetime. 
Instead, the surface of last scattering is a space-like hypersurface that occurs at some specific epoch of cosmic time when the ionization fraction drops to some suitable threshold level. 

If we want to estimate $\dot T_{\scriptscriptstyle CMB}$, the temperature drift of the CMB, then we need to think very differently. 
A more careful analysis of the surface of  last scattering requires working with the Saha equation to estimate the free electron fraction $x_e$ (see, for example,~\cite{last-scattering}):
\begin{equation}
{x_e^2\over 1-x_e} = {1\over n_H + n_p} \left(m_e k_B T\over2\pi \hbar^2 \right)^{3/2} \exp(-E_1/[k_BT]). 
\end{equation}
{Now,} $n_H + n_p= n_0 (1+z)^3$, and $T=T_0 (1+z)$, where $n_0$ and $T_0$ are measurable. (Here, $E_1$ is the ionization energy of the hydrogen atom.) 
By picking some specific free electron fraction, $x_0$, to characterize the surface of 
last scattering, one can then solve for $z_{\scriptscriptstyle{\mathrm{last\,scattering}}}$:
\begin{equation}
(1+z_{\scriptscriptstyle{\mathrm{last\,scattering}}})^{-3/2} \exp\left(-{E_1\over k_B T_0} \;{1\over1+z_{\scriptscriptstyle{\mathrm{last\,scattering}}}}\right) 
= n_0 \; {x_0^2\over 1-x_0} \left(m_e k_B T_0\over2\pi \hbar^2 \right)^{-3/2}.
\end{equation}
{Thence}
\begin{equation}
1+ z_{\scriptscriptstyle{\mathrm{last\,scattering}}} = 
{{2E_1\over 3k_B T_0}\over W\left(-{2E_1\over 3 k_BT_0} \; {2\pi \hbar^2\over m_e k_B T_0 } \;  n_0^{2/3} \left({x_0^2\over 1-x_0}\right)^{2/3}\right)}.
\end{equation}
{Here,} $W(x)$ is the Lambert $W$ function, defined by $W(x) \, e^{W(x)}= x$, which is a special function that is becoming increasingly important in both theoretical physics and \mbox{mathematics~\cite{Corless:1996,Valluri:2000,Valluri:2009,Boonserm:2013,Sonoda:2013a,Visser:2018-LW,Charters:2009ku}.}

The physics point here is that
 last scattering is a specific epoch in the evolution of the universe, a space-like hypersurface, with specific values for $z_{\scriptscriptstyle{\mathrm{last\,scattering}}}$, $T_{\scriptscriptstyle{\mathrm{last\,scattering}}}$, and $n_{\scriptscriptstyle{\mathrm{last\,scattering}}}$, 
not a physical time-like surface at a specific comoving position in the universe (that is, not a time-like hypersurface). The present-day value of the CMB temperature is, therefore, simply
\begin{equation}
T_{\scriptscriptstyle{\mathrm{CMB},0}}= {T_{\scriptscriptstyle{\mathrm{last\,scattering}}}\over 1+ z_{\scriptscriptstyle{\mathrm{last\,scattering}}}} = T_0.
\end{equation}
{If} we now project this into the future (or into the past), the CMB temperature will be
\begin{equation}
T_{\scriptscriptstyle{\mathrm{CMB}}}(t) = T_0 \;\; {a_0\over a(t)}.
\end{equation}
{This} is, of course, a very well-known result. What might not be so well understood is the physical understanding of \emph{{why}} this result cannot be reproduced via an RD analysis, which we hoped to have clarified. If we now differentiate around the present epoch, then
\begin{equation}
\dot T_{\scriptscriptstyle{\mathrm{CMB},0}}= - H_0 \;T_{\scriptscriptstyle{\mathrm{CMB},0}}. 
\end{equation}
{That} is, there is a drift in the temperature of the CMB, but it is appallingly small---one part in $10^{10}$ per year. 
For the near future, however, this effect is completely ruled out of being measured. 
Besides the quality in the CMB data available, we are still extremely far from the 10 significant digits.
Indeed, if one were to find a detectable drift in the temperature of the CMB, then this would be very difficult to reconcile with standard FLRW cosmology.

\subsubsection{Higher-Order RD}
\enlargethispage{20pt}
Based simply on dimensional analysis, all higher-order RDs are of the form:
\begin{equation}
    z^{(n)}_0 = \left.{d^n z \over d t^n}\right|_0 = \hbox{(dimensionless number)} \times H_0^n.
\end{equation}
{The} point is that the extra factors of $H_0$ will suppress all of these higher-order RD effects by an extra factor $[1/\hbox{(Hubble time)}]^n$, making them unobservably small---unless you commit to multi-millennia-long observational programs of the Loeb variety~\cite{Loeb:2022}. So, while the higher-order RDs may be theoretically important, their observational relevance is much less clear.

\subsection{Summary}

We have seen (above) a number of exact and perturbative results for the RD $\dot z(z)$. Perhaps the central point to take from this discussion is that for plausible cosmological models, the RD exhibits a local maximum at $z\approx 1$ and an ``accidental'' zero for $z\approx 2$. 
(The zero is ``accidental'' in that its location and very existence depend on the precise value of the $\Omega_{0i}$ parameters. For the hypothetical case $\langle w\rangle_0 > -1/3$, which is not a good model for our observed universe, for which $\langle w\rangle_0 \approx -0.685$, this zero would not exist.)
This observation is central when selecting possible candidates for observational study.

\section{Redshift Drift for Different Dark Energy Models}
\label{S:dark_energy_models}

Another important question regarding the future of precision cosmology is, 
\emph{{"Will RD data be able to differentiate dark energy models from each other?"}}
Following the lines of \cite{Balbi:2007fx} (see also \cite{Corasaniti:2007}), here, we present an analysis of different dark energy models, including $w_0$CDM, the linear model, CPL, BAZS, and a couple of interactive possibilities. 
Given the redshift evolution of the Hubble parameter $H(z)$ for each model, we are able to calculate the RD using the following relation \eqref{keyeq}:
\begin{equation}
    \label{h/h}
    \dot z = H_0 \left( 1+z - \frac{H(z)}{H_0} \right).
\end{equation}
{Before} presenting the results, let us first introduce each one of the models discussed here:\\

\noindent{\bf $w_0$CDM:} 
 (or simply $w$CDM): It is the simplest dark energy model, where the barotropic equation of state is given by $p = w_0 \rho$, with $w_0$ being a constant. In this case, as presented above, the Hubble parameter evolves as
\begin{equation}
\label{wcdm}
\left(\frac{H(z)}{H_0}\right)^2 =  \Omega_{r} (1+z)^{4} + \Omega_{m} (1+z)^{3} + \Omega_{DE} (1+z)^{3(1+w)}.
\end{equation}
{Figure} \ref{fig:DE1}a shows the results of the theoretically predicted RD signal \eqref{h/h} for different values of $w_0$.

\smallskip
\noindent{\bf{BAZS EoS:}}
The Barboza–Alcaniz–Zhu–Silva \cite{BAZS} equation of state is a three-parameter model for dark energy given by
\begin{equation}
    w = w_0 - w_b \;\dfrac{(1+z)^{-b} -1}{b}\;.
\end{equation}
{This} model includes the linear model as the limit of $b\to -1$ and CPL as the limit of $b\to 1$. Furthermore, taking the limit of $b\to 0$ gives us the logarithmic model. 
The Hubble parameter evolution, assuming $\Omega_r = \Omega_k =0$, is then given by~\cite{BAZS}
\begin{equation}
 \left(\frac{H(z)}{H_0}\right)^2 =  \Omega_m (1 + z)^3 + \Omega_{DE}\;(1 + z)^{3(1 + w_0 + \frac{w_b}{b})}  \;\exp\left[\frac{3 w_b}{b^2} \left(\frac{1}{(1+z)^b} -1\right)\right]
\end{equation}
The results for a fixed $w_0=-1$ and $w_b=-1$ for different values of the parameter $b$ are presented in Figure \ref{fig:DE1}b.

\smallskip
\noindent{\bf{Linear model:}}  
In this model, the barotropic equation of state is allowed to vary along the evolution of the universe in a linear way:
\begin{equation}
    w = w_0 +w_b\;z\;.
\end{equation}
{The} two free variables $w_0$ and $w_b$ are constants, and the Hubble parameter evolution is given by
\begin{equation}
 \left(\frac{H(z)}{H_0}\right)^2 =  \Omega_m (1 + z)^3 + \Omega_{DE}\;(1 + z)^{3\PC{1 + w_0 - w_b}}  \;\exp\PC{3 w_b\; z} 
\end{equation}
{Figures} \ref{fig:DE1}c,d show how the RD data changes when keeping, respectively, $w_0$ and $w_b$ fixed while varying the other variable.

\smallskip
\noindent{\bf {CPL:}}
Named after Chevallier, Polarski, and Linder~\cite{CP,L}, this two-parameter equation of state guarantees, when compared to the linear model, a bound on how much $w$ can grow as we go back in time. 
\begin{equation}
    w = w_0 + w_b \;\frac{z}{1+z}
\end{equation}
{Here,} $w_0$ represents the value of $w$ at the present moment, while $w_0 + w_b$ represents its value at the asymptotic past. In this case, we have
\begin{equation}
\label{CPL}
\left(\frac{H(z)}{H_0}\right)^2 = \Omega_{r} (1+z)^{4} + \Omega_{m} (1+z)^{3} + \Omega_{DE} (1+z)^{3(1+w_0 +w_b)}\exp\left(\frac{-3w_b \;z}{1+z}\right).
\end{equation}
{The} results for the expected RD signal for distinct values of $w_b$ and $w_0$ can be found in \mbox{Figures \ref{fig:DE1}e,f,} respectively.

{
\begin{figure}[H] 
\begin{subfigure}{0.48\textwidth}
\includegraphics[width=\linewidth]{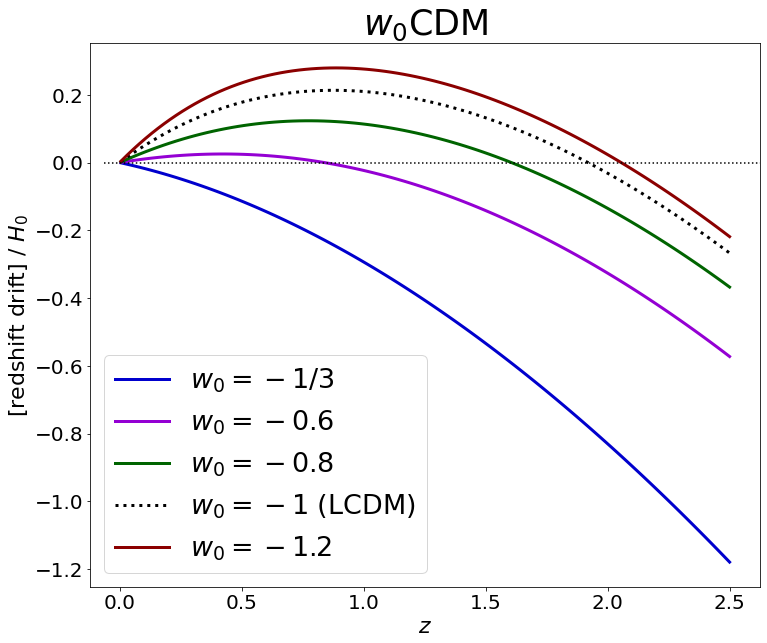}
\caption{$w_0$CDM for distinct values of $w_0$.} \label{fig:DE1a}
\end{subfigure}~~
\begin{subfigure}{0.48\textwidth}
\includegraphics[width=\linewidth]{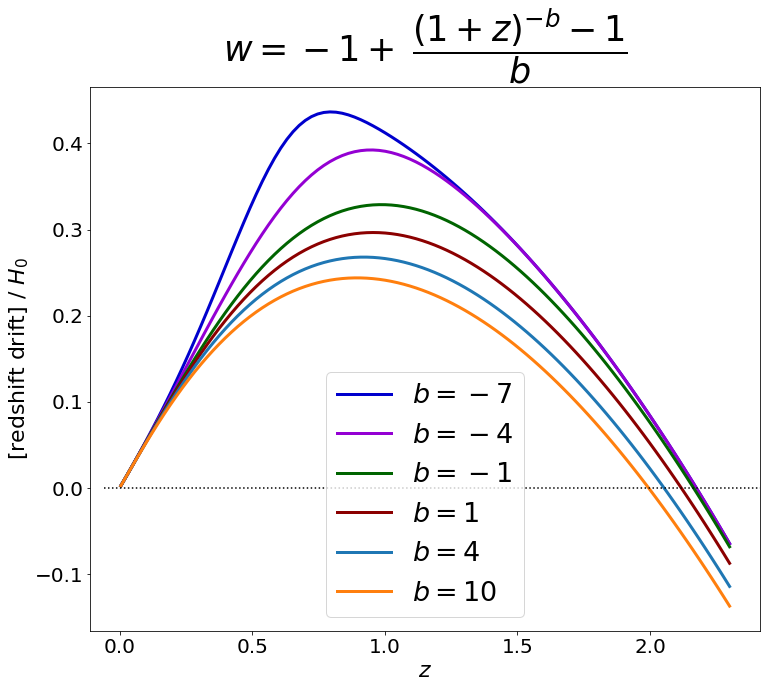}
\caption{BAZS EoS with fixed $w_0= w_b=-1$.} \label{fig:DE1b}
\end{subfigure}

\medskip
\begin{subfigure}{0.48\textwidth}
\includegraphics[width=\linewidth]{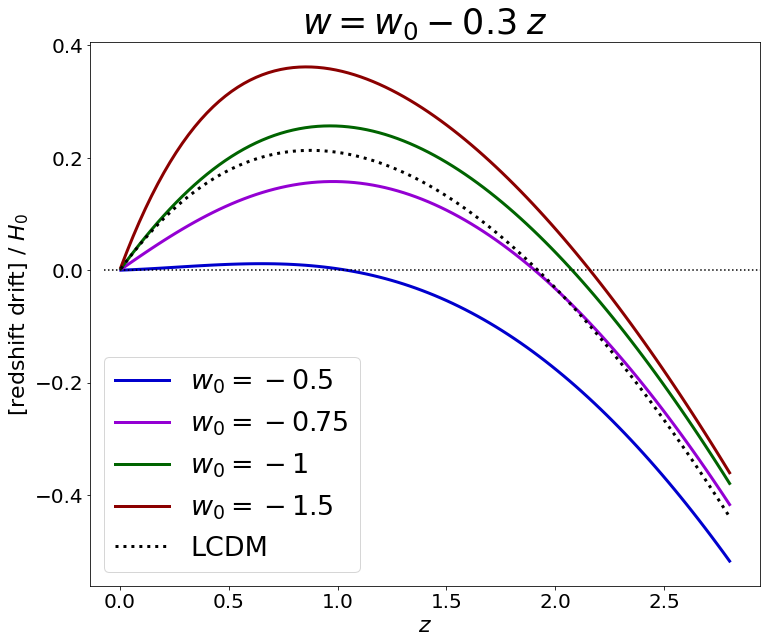}
\caption{Linear model with fixed $w_b=-0.3$.} \label{fig:DE1c}
\end{subfigure}~~
\begin{subfigure}{0.48\textwidth}
\includegraphics[width=\linewidth]{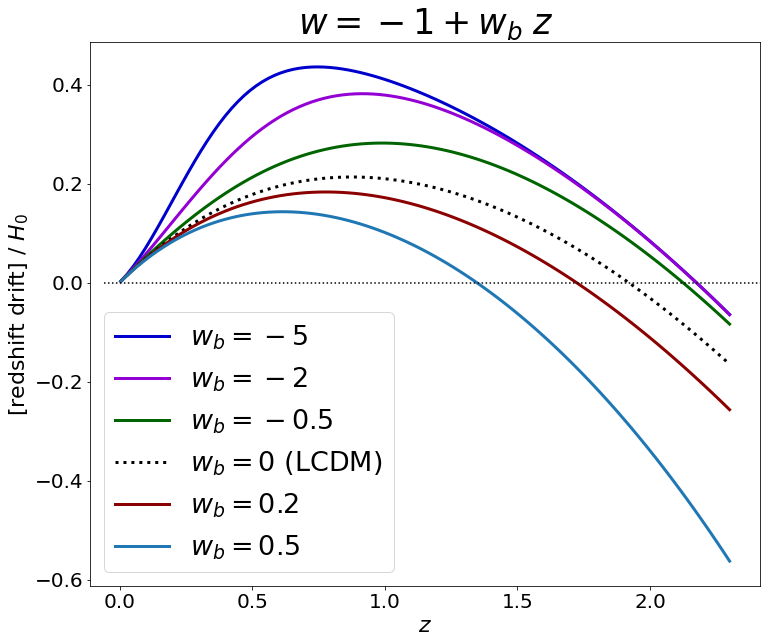}
\caption{Linear model with fixed $w_0=-1$.} \label{fig:DE1d}
\end{subfigure}

\medskip
\begin{subfigure}{0.48\textwidth}
\includegraphics[width=\linewidth]{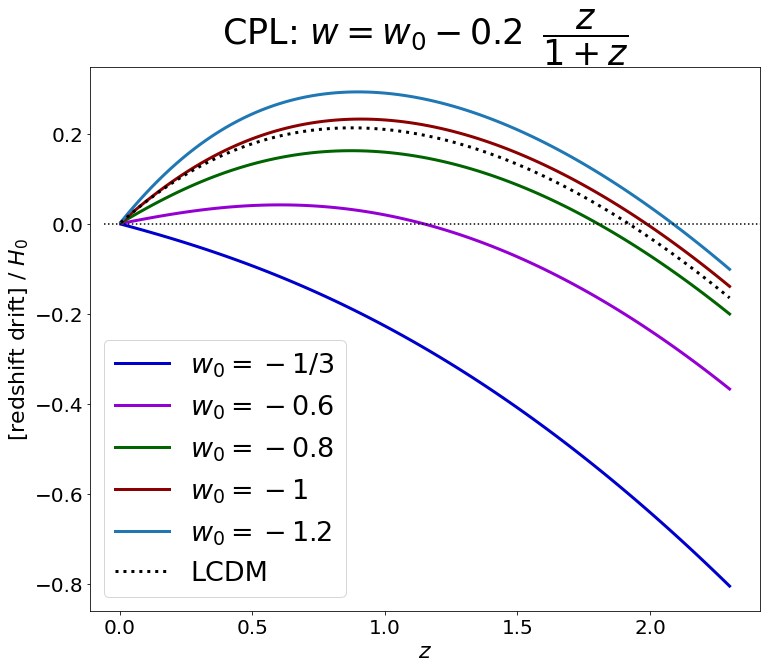}
\caption{CPL model with fixed $w_b=-0.2$.} \label{fig:DE1e}
\end{subfigure}~~
\begin{subfigure}{0.48\textwidth}
\includegraphics[width=\linewidth]{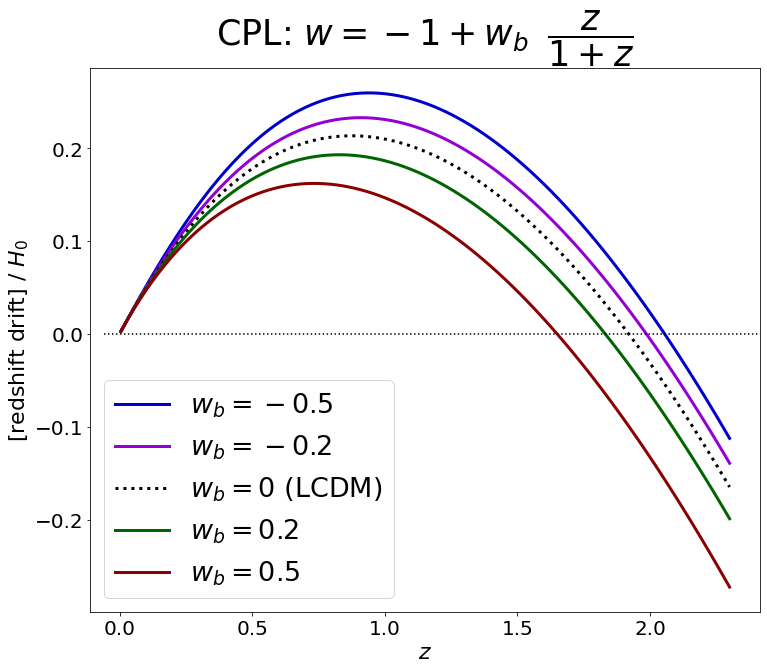}
\caption{CPL model with fixed $w_0=-1$.} \label{fig:DE1f}
\end{subfigure}
\vspace{6pt}
\caption{Redshift drift for different dark energy models.} \label{fig:DE1}
\end{figure}
}

\begin{figure}[H]
\begin{subfigure}{0.48\textwidth}
\includegraphics[width=\linewidth]{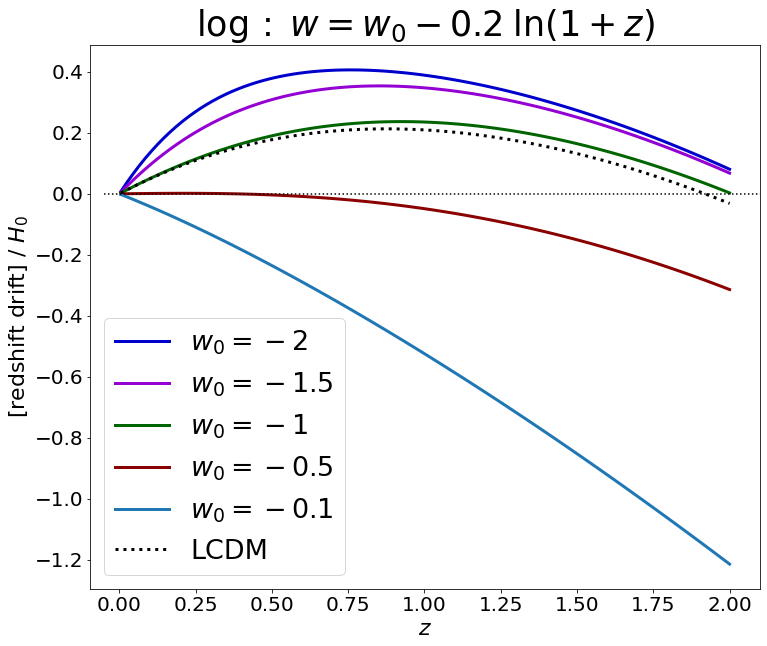}
\caption{Logarithmic model with fixed $w_b=-0.2$.} \label{fig:DE2a}
\end{subfigure}~~
\begin{subfigure}{0.48\textwidth}
\includegraphics[width=\linewidth]{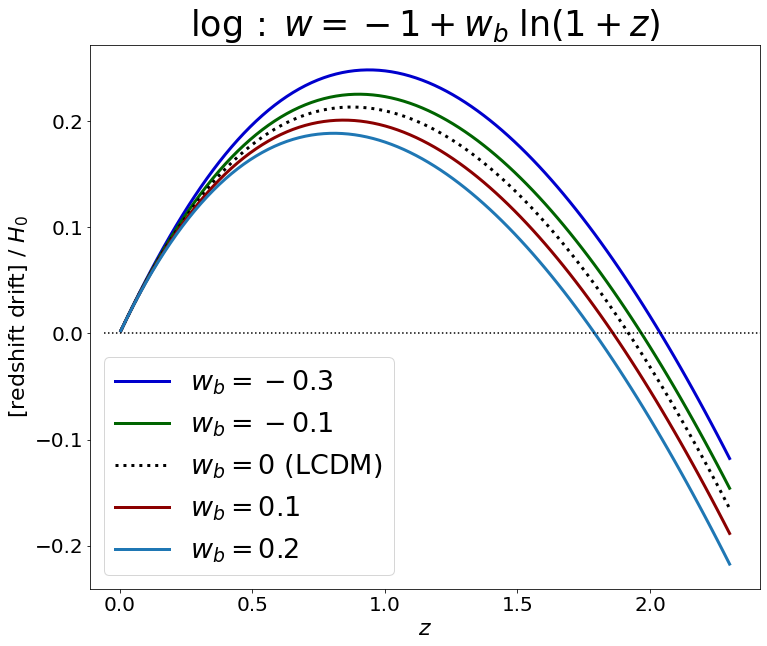}
\caption{Logarithmic model with fixed $w_b=-0.2$.} \label{fig:DE2b}
\end{subfigure}
\vspace{6pt}
\caption{Redshift drift for the logarithmic model.} \label{fig:DE2}
\end{figure}

\smallskip
\noindent{\bf{Logarithmic evolution:}}
This was first introduced by Efstathiou~\cite{Efs}, who suggested a logarithmic evolution of $w(z)$ into the asymptotic past:
\begin{equation}
    w = w_0 + w_b \;\ln(1+z)\;.
\end{equation}
{The} results for different values of $w_b$ and $w_0$ are presented in Figures \ref{fig:DE2}a,b, and the Hubble parameter evolution is given by
\begin{equation}
 \left(\frac{H(z)}{H_0}\right)^2 =  \Omega_m (1 + z)^3 + \Omega_{DE}\;(1 + z)^{3\PC{1 + w_0 +\frac{w_b}{2}\ln(1+z)}}\;.  
\end{equation}

\vspace{-12pt}

\smallskip
\noindent{\bf{Interactive dark energy models:}}
Another possible scenario is to consider interactive models, where dark energy and dark matter exchange energy via an interaction term, $Q$. In this case, the energy conservation equations in FLRW give us
\begin{eqnarray}
    \dot{\rho}_c + 3H\rho_c &=& Q\;,\\
    \dot{\rho}_{DE} + 3H(\rho_{DE} + p_{DE}) &=& -Q\;.
\end{eqnarray}
\textls[-15]{{Here,} $\rho_c$ and $\rho_{DE}$ stand for the energy density of cold dark matter and dark energy, respectively.}

\begin{figure}[H]

\begin{adjustwidth}{-\extralength}{0cm}
\centering 
\begin{subfigure}{0.48\textwidth}
\includegraphics[width=\linewidth]{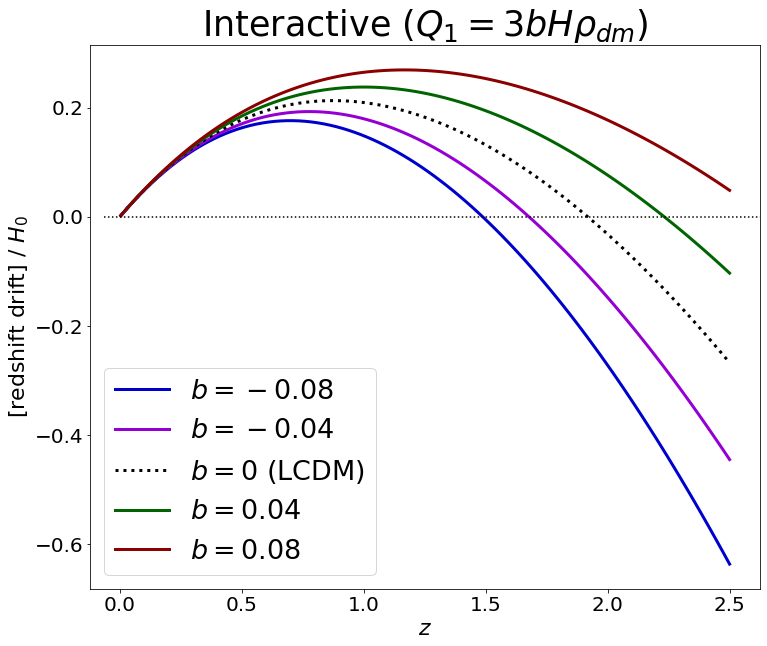}
\caption{\mbox{Interactive $Q_1$ model for varying coupling parameters.}} \label{fig:DE2c}
\end{subfigure}~~~~~~~~~~~~~~~~~~~
\begin{subfigure}{0.48\textwidth}
\includegraphics[width=\linewidth]{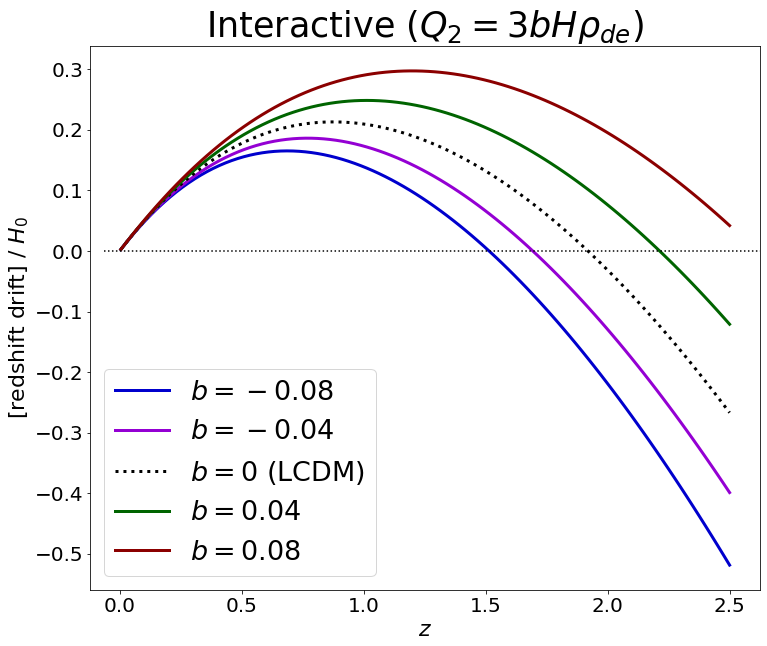}
\caption{\mbox{Interactive $Q_2$ model for varying coupling parameters.}} \label{fig:DE2d}
\end{subfigure}
\vspace{6pt}
\end{adjustwidth}
\caption{{Redshift} 
 drift for interactive dark energy models $Q_1$ and $Q_2$.} \label{fig:DE21}
\end{figure}

There are a few common possibilities for the interaction term. Here, we will present the RD results for $Q_1 = 3b H \rho_{DE}$ and $Q_2 = 3b H\rho_c$, where $b$ represents a dimensionless coupling parameter. 
The Hubble parameter evolution for $Q_1$ is given by~\cite{int1, int2}
\begin{equation}
   \left(\frac{H(z)}{H_0}\right)^2 =  \Omega_m (1+z)^3 + \frac{\Omega_{DE}}{(w_0 +b)} \bigl[ b(1+z)^3 +w_0 (1+z)^{3(1+b +w_0)} \bigr]
\end{equation}
and the redshift results can be found in Figure \ref{fig:DE21}a. For $Q_2$, on the other hand, the Hubble parameter behaves as~\cite{int1, int2}
\vspace{-12pt}
\begin{adjustwidth}{-\extralength}{0cm}
\begin{equation}
\qquad\qquad
   \left(\frac{H(z)}{H_0}\right)^2 =  \Omega_{DE}(1+z)^{3(1+w_0)} +\; \Omega_b (1+z)^3 + \frac{\Omega_{c}}{(w_0 +b)}
   \bigl[ b(1+z)^{3(1+w_0)} + w_0 (1+z)^{3(1-b)} \bigr]\;.
\end{equation}
\end{adjustwidth}
{Note} that $\Omega_m = \Omega_b + \Omega_c$, with this separation being necessary, given that baryons clearly do not interact with either dark energy or cold dark matter. The results for the different values of $b$ can be found for $Q_2$ in Figure \ref{fig:DE21}b.

\vspace{-3pt}

\subsection*{Can RD Data Distinguish Different Models?}
\label{S:DE_discussion}
In view of the last decade, with the increasing flow of new data and different dark energy models being proposed, a natural question to ask is whether different models can be discerned from each other and which data can be useful in this challenging task~\cite{int1, int2, DET1, DET2, DET3, DET4, DET5}.

As must now be a full consensus, no particular data \emph{{alone}} can resolve that question. Our goal here is to emphasize that, as great as the future RD data might be, the result is still the same---RD data alone cannot separate one model from another. 

Our approach to show this will be to compare the theoretically predicted signals for distinct models.
Given that different models may have from one to three free variables to fit the data, it may not come as a surprise that very different models might present curves that---including future error bars---will be simply indistinguishable from each other. Furthermore, given how clearly this can be shown analytically, we have refrained from performing an MCMC analysis due to the lack of need.

\textls[-15]{In all of the comparative analysis presented here, the results are shown for $z\leq 2.5$ and higher redshifts separately in order to present the curves with greater detail. 
Additionally, given that real RD data are not yet available, we have not included any error bars. These have been included in previous works using the results from Monte Carlo {simulations}~
\cite{codex}, where the error on the spectroscopic velocity shift of CODEX was predicted. We have decided, however, to keep those out, given that our point can be proven even without their inclusion. The curves were obtained with optimization algorithms in Python.}

\begin{figure}[H]
\includegraphics[width=0.6\linewidth]{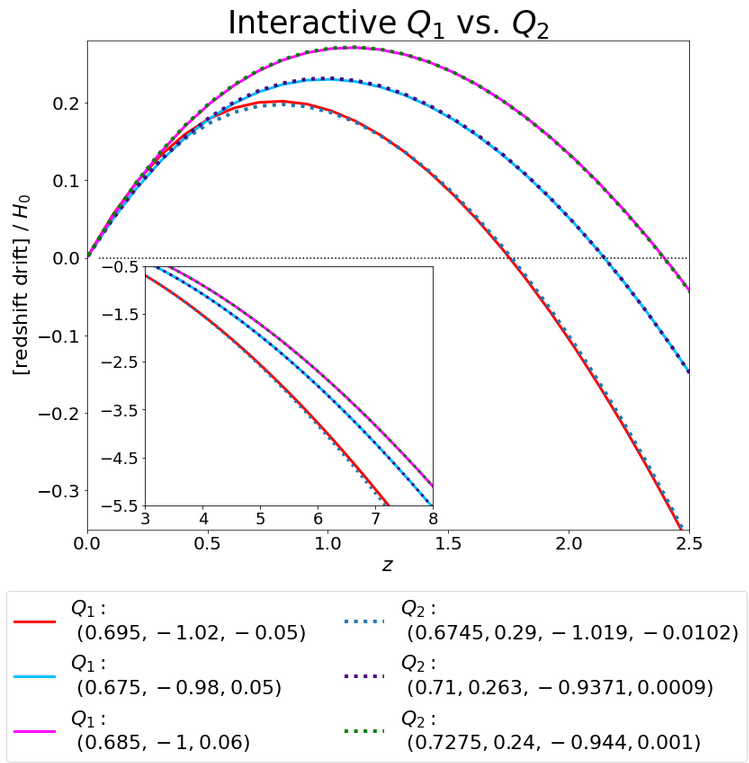}
\caption{Comparison between the interactive models $Q_1 = 3bH\rho_{DE}$ and $Q_2 = 3bH\rho_c$.} \label{fig:vs_int}
\end{figure}

Let us start by looking at the results of a comparison between the interactive models $Q_1$ and $Q_2$ presented in Figure \ref{fig:vs_int}. In this graph, we input three curves for the interactive model $Q_1$ and found the best-fit parameters for the model $Q_2$ in order to reproduce the same curves. The bounds assumed for the fitting parameters of $Q_2$ were given by $\Omega_{DE} \in [0.65, 0.73]$ and $\Omega_c \in [0.24, 0.29]$, while $w_0 \in [-2, 0]$ and $b \in [-1, 1]$.
Furthermore, we have applied the following constraints: i) $\Omega_{DE} + \Omega_c \approx 0.9545$, in order to place a bound on the amount of baryonic matter; ii) $b\neq 0$, so as to guarantee that we have an interactive model. Both constraints had a tolerance of $10^{-3}$. As it is clear from the figure, some of these lines almost perfectly superpose to each other. 

\begin{figure}[H]
\includegraphics[width=0.6\linewidth]{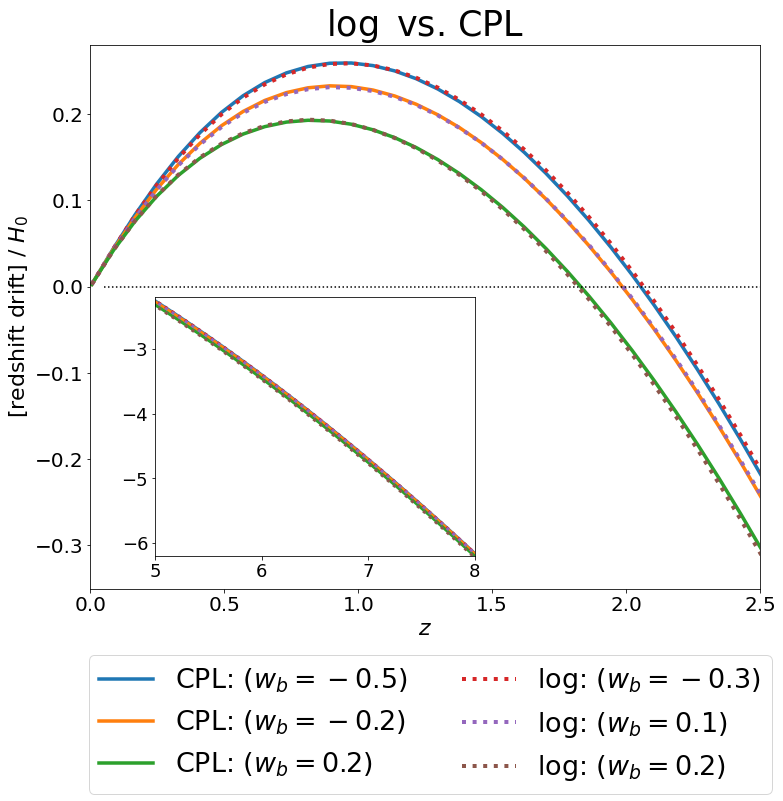}
\caption{Comparison between the logarithmic and CPL models. Here, we have kept $w_0 =-1$ for both models. The results for different values of $w_b$ are presented.} \label{fig:vs_P2_CPL}
\end{figure}

\begin{figure}[H]
\includegraphics[width=0.75\linewidth]{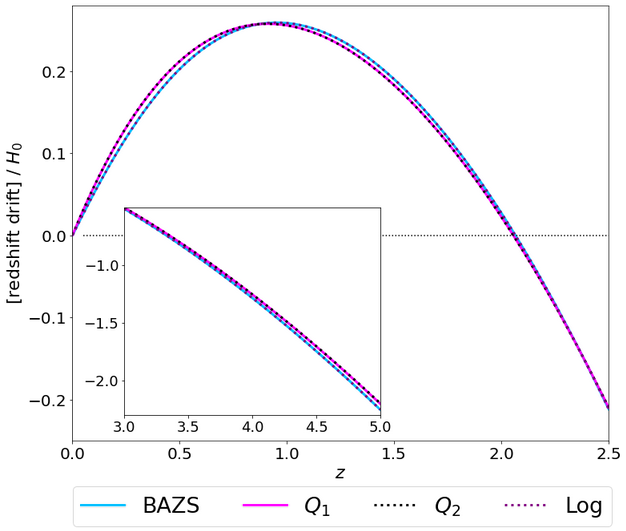}
\caption{Generated logarithmic curve and the best-fit curves found for the BAZS and interactive models $Q_1$ and $Q_2$.} \label{fig:vs_4}
\end{figure}

Figure \ref{fig:vs_P2_CPL} presents a similar analysis comparing the logarithmic and CPL models. Here, the curves of CPL were fit to a generated logarithmic model curve with fixed \mbox{$\Omega_{DE}=0.685$} and fixed $w_0=-1$, while allowing $w_b \in [-1, 1]$ to vary.
Finally, in Figure \ref{fig:vs_4}, we present a similar analysis fitting the BAZS, $Q_1$, and $Q_2$ models to a generated logarithmic curve. The bounds on $\Omega_{DE}$, $w_0$, and $w_b$ were the same as those applied in Figure \ref{fig:vs_int}. The input logarithmic curve and the best-fit parameters obtained {can be seen in the following} 

\begin{center}
\begin{tabularx}{\textwidth}{ |c|C|} 
 \hline
$\log$: & $\Omega_{DE} = 0.685$, $w_0 = -1$, $w_b = -0.4$\\
BAZS: & $\Omega_{DE} = 0.6816$, $b = 0.0139$, $w_0 = -1.1099$\\
$Q_1$: &$\Omega_{DE} = 0.6850$, $b = 0.01$,  $w_0 = -0.9996$, $w_b = -0.4019$\\
$Q_2$: & $\Omega_{DE} = 0.6894$, $\Omega_{c} = 0.2565$, $b = 0.001$, $w_0 = -1.0989$\\
\hline
\end{tabularx}
\end{center}

The point to be taken from these analyses should be very clear: independently of how precise future redshift drift data might be, different dark energy models cannot be distinguished from this type of data alone. The scenario improves (meaning that no good fits between different models are available) when the bounds on the free parameters are tightly constrained by other data sources, especially when higher redshift $(z >10)$ points are included in the analysis.

\vspace{-9pt}

\vspace{-3pt}

\section{Redshift Drift for Other Auxiliary Variables}

The current status of cosmography is quite subtle \cite{Aviles:2012, Dunsby:2015, Busti:2015, Lusso:2020, Hu:2022, Yang:2020, Capozziello:2020,Lobo:2020}. While it is certainly well appreciated that the luminosity distance expansion in terms of $z$ does not converge, there is still considerable debate on which \emph{{auxiliary function}} is the best replacement for it. The discussions regarding this subject seem to not have a final conclusion, and debates from auxiliary functions that ``produce'' (or display) tensions with $\Lambda$CDM data have been raised in the past \cite{Busti:2015, Lusso:2020, Yang:2020}.
However, since discussing this topic could lead to a manuscript in its own right, we will focus on simply enumerating the most well-known of the so-called \emph{{auxiliary variables}}, including the pioneer, $y$, followed by their respective RDs---for which we shall perform both cosmographic and cosmodynamic analyses. We will present a more detailed discussion of the $y$ redshift case, followed by a table with the main results for the other variables.

For some background, recall the standard cosmographic result
\begin{equation}
    H(z) = H_0 \left\{ 1+ (1+q_0) z + {1 \over 2} (j_0-q_0^2) z^2 +\mathcal{O}(z^3) \right\},
\end{equation}
which we shall subsequently rephrase in terms of the new auxiliary variables.

\subsection{Redshift Drift in Terms of $y$}
\label{S:y-redshift}
In reference~\cite{Lobo:2020}, working in terms of the $y$-redshift 
\begin{equation}
1-y = {a\over a_0} = {1\over 1+z},
\end{equation}
three of the current authors demonstrated that
\begin{equation}
\dot y = (1-y)H_0 - (1-y)^2 H(y).
\label{ydot}
\end{equation}
{By expanding} $H(y)$ and keeping those terms only up to the second order in the cosmographic expansion, we have
\begin{equation}
    H(y) = H_0 \left\{ 1+ (1+q_0) y + \Big[1+q_0 +{1 \over 2}(j_0-q_0^2)\Big] y^2 +\mathcal{O}(y^3) \right\}.
\end{equation}
{Thence,} for the RD,
\begin{equation}
\label{RD y}
    \dot{y} = - H_0\left\{ q_0\; y - \frac{1}{2}\;\PC{  2q_0 +  q_0^2 - j_0 } y^2
    +\mathcal{O}(y^3) \right\}.
\end{equation}
{In order to} proceed to a cosmodynamic analysis, we first write the general model-independent result:
\begin{equation}
\dot y  = (1-y) \left\{ 1 -(1-y)\sqrt{\Omega(y)}\right\}H_0;  \qquad \qquad  (\Omega_0\equiv1). 
\end{equation}
{This} is still exact in any FLRW spacetime.

So, assuming the generic matter model $\{\Omega_{oi},w_i\}$ for the EOS, we obtain the following result:
\begin{equation}
\dot y = H_0 (1-y) \left\{ 1-(1-y) \sqrt{ \sum_{i=1}^N \Omega_{0i} (1-y)^{-3-3w_i} } \right\};
\qquad\qquad
\sum_{i=1}^N \Omega_{0i} = 1.
\label{rd y omega}
\end{equation}
If we now Taylor expand for a small $y$, that is, $|y|<1$, then, for the Hubble parameter, one has\vspace{-12pt}
\begin{adjustwidth}{-\extralength}{0cm}
\centering 
\begin{equation}
\qquad\qquad\qquad
H(y) = {H_0} \PR{ 1 + {1\over2}\left(\sum_{i=0}^N (3+3w_i) \Omega_{0i}\right) y + \mathcal{O}(y^2)} = {H_0} \PR{ 1 + {3\over2} \left( 1+\langle w\rangle_0  \right) y + \mathcal{O}(y^2)}\;.
\end{equation}
\end{adjustwidth}
{When} inserted into the general expression for $\dot y$, one finds
\begin{equation}
\dot y =- {H_0 y\over 2} \left\{ 1 + 3\langle w\rangle_0 + \mathcal{O}(y) \right\}.
\end{equation}
\mbox{This} is compatible with our earlier lowest-order result for $\dot z$. Note, however, that since it is now expressed in terms of $y={z\over1+z}$, the range of applicability for the Taylor expansion is much wider than in the $z$-redshift case. In particular, the peak at $z\approx 1$ now occurs at $y\approx 1/2$, and the zero at $z\approx 2$ now occurs at $y\approx {2/3}$.
Indeed, $|y|<1$ covers the entire physical region $z\in[0,\infty)$. 
If we go to one order higher in $y$, we get
\begin{equation}
\dot y =- {H_0 y\over 2} \left\{ 1 + 3\langle w\rangle_0 
+{1\over4} \left[18 \langle w^2\rangle_0 
-9 \langle w\rangle_0^2  -1 \right]y +
\mathcal{O}(y^2) \right\}.
\label{rd y omegabar}
\end{equation}
{Let} us now consider some more specific models.

\paragraph*{\bf{One-component model: }}

When one component dominates over all the others, we simply have
\begin{equation}
\dot y = H_0 (1-y) \left\{ 1-(1-y)  (1-y)^{-3(1+w)/2}  \right\}.
\end{equation}
{Just} as before, the $y-$RD is positive for $w<-1/3$, zero for $w = -1/3$, and negative for $w>-1/3$. For the special case of de Sitter space, $(w=-1)$, this further simplifies to
\begin{equation}
\dot y = H_0 \; y (1-y); \qquad (\text{for} \; w=-1)\;.
\end{equation}

\paragraph*{\bf{$\Lambda$CDM (general four-component version): }}

\begin{figure}[H]
    \includegraphics[scale=0.5]{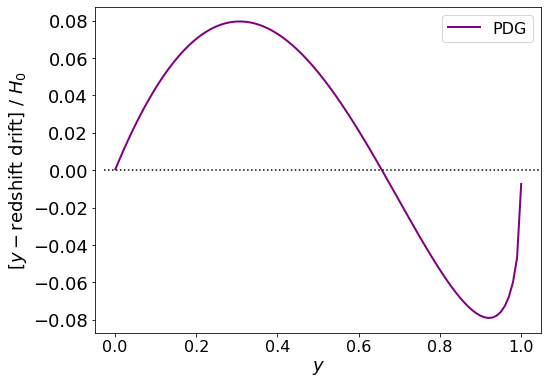}
    \caption{{$y-$Redshift} 
 drift (per $H_0$) for four-component $\Lambda$-CDM, estimated from the PDG data.}
    \label{F:ydriftPDG}
\end{figure}

For the four-component $\Lambda$CDM model ($\Lambda$ + curvature+ dust + radiation), the $y-$RD takes the form
\begin{equation}
\dot y = H_0 (1-y) \left\{ 1-(1-y) \sqrt{ \Omega_\Lambda + \Omega_k(1-y)^{-2}  + \Omega_m(1-y)^{-3} +\Omega_r (1-y)^{-4}}\right\},
\end{equation}
with
\begin{equation}
\Omega_\Lambda + \Omega_k  
+ \Omega_m  + \Omega_r =1.
\end{equation}
{Making} use, again, of the latest PDG data, we find that this curve has three zeros at $y\in\{0,0.657,1\}$. Note that $y \approx 0.657$ corresponds to $z\approx 1.918$ (the same value obtained before, as expected). See Figure \ref{F:ydriftPDG}.

\vspace{-6pt}

\paragraph*{\bf{$\Lambda$CDM (two-component version): }}

\begin{figure}[H]
    \includegraphics[scale=0.5]{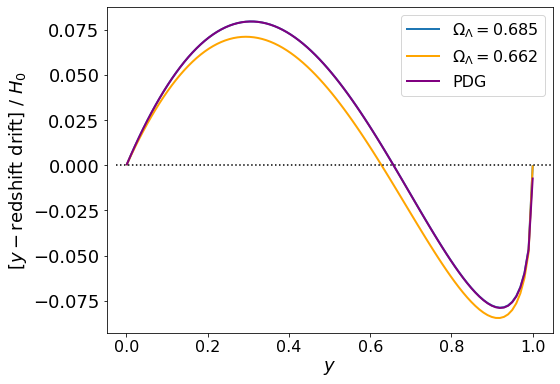}
    \caption{$y-$RD magnitude (per $H_0$) for two-component $\Lambda$-CDM for different values of $\Omega_{\Lambda}$. As before, we have taken the values for $\Omega_{m} = 0.315\pm 0.007$ given by the Planck2018 data  release~\cite{Planck2018} (blue curve) and $\Omega_m = 0.338 \pm 0.018$ from Pantheon+~\cite{Pantheon+} (yellow curve) plus the values previously given by the PDG data (purple curve). Again, since both Planck (assuming $\Omega_k = \Omega_r = 0$) and PDG have exactly the same value for $\Omega_{\Lambda}$, the two curves appear to exactly coincide.}
    \label{fig:y-RD}
\end{figure}

For the two-component $\Lambda$CDM model ($\Lambda$ + dust), we then have
\begin{equation}
\dot y = H_0 (1-y) \left\{ 1-(1-y) \sqrt{\Omega_\Lambda + (1-\Omega_\Lambda) (1-y)^{-3} } \right\}.
\end{equation}
{In} Figure \ref{fig:y-RD}, we superimpose two curves based on the Planck2018 and Pantheon+ data.
Again, for $\Omega_\Lambda \approx 0.685$, there are three zeros of $\dot y(y)$ at $y\in\{0,0.657,1\}$, a maximum $\dot y_{max} = 0.0795$ at $y_{max}=0.307$,
and a minimum $\dot y_{min} = -0.0788$ at $y_{min}=0.920$.
This is qualitatively and quantitatively in agreement with what we saw happening in the $z$-redshift. 

In summary, the $y$ redshift maps the entire infinite range $z\in [0,\infty)$ into the finite interval $y\in[0,1)$.
Because of this, the $y$ redshift is both theoretically preferable and pragmatically useful---in particular, $y_\mathrm{CMB} \approx 0.999$---so that one can, in principle, backtrack to decoupling without having the figures expand off the page. When working in terms of the $y$ redshift, the qualitative features of the RD $\dot y(y)$ closely parallel those of $\dot z(z)$.

\vspace{-3pt}

\subsection{Table of Expressions for Other Auxiliary Variables}
\label{S:other_auxiliary_variables}
In the table below, we have exposed the relevant expressions for the RD for some of the most well-known auxiliary variables. These results are the direct analog of equations \eqref{ydot}--\eqref{RD y}, \eqref{rd y omega}, and \eqref{rd y omegabar} for different {variables.}

\begin{center}
\renewcommand{\arraystretch}{1.2}
\renewcommand{\aboverulesep}{.001pt}
\renewcommand{\belowrulesep}{.001pt}
\begin{tabularx}{\textwidth}{ |c|C|} 
 \hline

  \small{\textbf{{Variable:}}} & $H(y) = H_0 \left\{ 1+ (1+q_0) y + \Big[1+q_0 +{1 \over 2}(j_0-q_0^2)\Big] y^2 \right\}$ \\

$y = \frac{z}{1+z}$ &   $\dot y = (1-y)H_0 - (1-y)^2 H(y)$ \\

 $z = \frac{y}{1-y}$ &   $\dot{y} = - H_0\left\{ q_0\; y - \frac{1}{2}\;\PC{  2q_0 +  q_0^2 - j_0 } y^2
    \right\}$ 
  \\

   \small{\textbf{{Convergence radius:}}} & $\dot y = H_0 (1-y) \left\{ 1-(1-y) \sqrt{ \sum_{i=1}^N \Omega_{0i} (1-y)^{-3-3w_i} } \right\}$ \\

    $|y| < 1 $     &  $\dot y =- {1 \over 2} H_0 \left\{ y + 3\langle w\rangle_0 
+{1\over4} \left[18 \langle w^2\rangle_0 
-9 \langle w\rangle_0^2  -1 \right]y^2  \right\}$ \\

 \hline

 \small{\textbf{Variable:}} & $H(y_1) = H_0 \left\{ 1+ (1+q_0) y_1 + {1 \over 2} (1+q_0-q_0^2+j_0) y_1^2  \right\}$ \\
$y_1 = \ln(1+z)$ &   $\dot{y}_1 = H_0 -e^{-y_1}\; H(y_1)$ \\
 $z = e^{y_1} -1$ &   $\dot{y}_1 = - H_0\left\{ q_0\; y_1 - \frac{1}{2}\;\PC{  q_0 +  q_0^2 - j_0 } y_1^2
     \right\}$ 
  \\

   \small{\textbf{{Convergence radius:}}} & $\dot{y}_1 = H_0 \left\{1 -e^{-y_1}\;\sqrt{\sum_{i=1}^N \Omega_{0i} \; e^{3 y_1 (1+w_i)}}\;\right\}$ \\

    $|y_1| < \infty $     &  $\dot{y}_1 = - {1\over2}H_0\left\{ \left(1+3\langle w\rangle_0 \right)\; y_1 
    + \left({1+6\langle w\rangle_0 +18\langle w^2\rangle_0-9 \langle w\rangle_0^2\over 4}\right)\; y_1^2
     \right\}$ \\

 \hline

  \small{\textbf{{Variable:}}} & $H(y_2) = H_0 \left\{ 1+ (1+q_0) y_2 + {1 \over 2} (j_0-q_0^2) y_2^2  \right\}$  \\

    $y_2 = \arctan(z)$ &    $\dot{y}_2 = \cos^2(y_2) \left\{ [1+\tan(y_2)]\; H_0 - \;H(y_2) \right\}$
   \\

  $z=\tan(y_2)$ & $\dot{y}_2 = - H_0\left\{ q_0 \;y_2- \frac{1}{2}\;(-j_0 + q_0^2)\; y_2^2
   \right\}$ \\

     \small{\textbf{{Convergence radius:}}} &  $\dot{y}_2 = H_0\; \cos^2(y_2) \left\{[1+\tan(y_2)] - 
   \sqrt{\sum_{i=1}^N \Omega_{0i} \; [1+\tan(y_2)]^{3 (1+w_i)}}\;\right\}$\\

   $ |y_2| < \frac{\pi}{4} $ & $\dot{y}_2 = - {1\over2}H_0\left\{ \left(1+3\langle w\rangle_0 \right)\; y_2 
    + \left({3+12\langle w\rangle_0 +18\langle w^2\rangle_0-9 \langle w\rangle_0^2\over 4}\right)\; y_2^2
     \right\}$ \\

 \hline

  \small{\textbf{{Variable:}}} & $H(y_3) = H_0 \left\{ 1+ (1+q_0) y_3 + {1 \over 2} (2+2q_0-q_0^2+j_0) y_3^2  \right\}$   \\

  $y_3 = \arctan(\frac{z}{1+z})$ &  $\dot{y}_3 = \cos^2(y_3) \{1-\tan(y_3)\}\; H_0 - \{1-\sin(2y_3)\}\;H(y_3)$
   \\

  $z= {\tan(y_3)\over 1-\tan(y_3)}$ & $\dot{y}_3 = - H_0\left\{ q_0 \;y_3+ \frac{1}{2}\;(2q_0-q_0^2+j_0)\; y_3^2
    \right\}$ \\

 \small{\textbf{{Convergence radius:}}} & $\dot{y}_3 = H_0\;  \bigl\{ \cos^2(y_3)[1 -\tan(y_3)] \qquad \qquad \qquad \qquad\quad$ \\

   $ |y_3|<\frac{\pi}{4} $ & 
   $\qquad \qquad - [1-\sin(2y_3)]  \sqrt{\sum_{i=1}^N \Omega_{0i} \; [1-\tan(y_3)]^{-3 (1+w_i)}}\bigr\}$ \\

    & $\dot{y}_3 = - {1\over2}H_0\left\{ \left(1+3\langle w\rangle_0 \right)\; y_3 
    - \left({1-18\langle w^2\rangle_0+9 \langle w\rangle_0^2\over 4}\right)\; y_3^2
    \right\}$ \\

 \hline
\end{tabularx}
\end{center}

\subsection{Summary}
The key point for all of these alternate auxiliary variables is that, at low redshift,
\begin{equation}
    y_i = z + \mathcal{O}(z^2).
\end{equation}
{Consequently,} for the lowest-level cosmographic expansion, we always have
\begin{equation}
    \dot{y}_i = - H_0\; q_0 \left\{ y_i +\mathcal{O}(y_i^2)\right\}.
\end{equation}
{For} our generic matter model $\{\Omega_{0i}, w_i\}$, for the lowest-level cosmodynamic expansion, we always have
\begin{equation}
\label{RD yi}
    \dot{y}_i = - {1\over2}H_0\left(1+3\langle w\rangle_0 \right)\; 
    \left\{  y_i +\mathcal{O}(y_i^2) \right\}.
\end{equation}
{So,} while the choice of auxiliary redshift variable $y_i$ might affect the specific range of physical interest and the precise location of any peaks or zeros in the RD, there are, nevertheless, strong qualitative similarities.

Note the RD peak at $z \approx 1$ corresponds to 
$y \approx 1/2$, $y_1 \approx \ln(2) \approx 0.693$, $y_2 \approx \arctan(1) \approx 0.785$, $y_3 \approx \arctan(1/2) \approx 0.464$,
all of which are safely less than 1. 
The zero at the RD occurs at $z \approx 2$, which corresponds to 
$y \approx 2/3$, $y_1 \approx \ln(3) \approx 1.099$, $y_2 \approx \arctan(2) \approx 1.107$, $y_3 \approx \arctan(2/3) \approx 0.588$, all of which are inside their respective convergence radii. 
Observe that the appropriate radii of convergence are 1, $\infty$, and \mbox{$\pi/4\approx 0.78539$. }

\section{Discussion and Conclusions}
\label{S:Conclusions}

In this article, we have extensively discussed the RD at both the cosmographic and cosmodynamic levels, focusing primarily on the cosmodynamic aspects. 
At the purely cosmographic level, in terms of suitable redshift variables $\{z, y, y_1\}$,
\begin{equation}
a/a_0= {1\over1+z} = 1-y = e^{-y_1},
\end{equation}
and the key formulae can be written as
\begin{equation}
    \dot z = (1+z) H_0 - H(z).
\end{equation}
\begin{equation}
\dot y = (1-y)\left\{ H_0 - (1-y) H(y)\right\}.
\end{equation}
\begin{equation}
   \dot{y}_1 = H_0 -e^{-y_1}\; H(y_1) \;.
\end{equation}
{At} the cosmodynamic level, when using the Friedmann equations only and when suitably defining the $\Omega$ parameter to include the effects of spatial curvature, one has $H = H_0 \sqrt{\Omega}$.
By assuming, initially, a generical model, which is then replaced by the specific cosmological scenario for which 
\begin{equation}
\Omega = \sum_{i=1}^N \Omega_{0i} \;(1+z)^{3(1+w_i)},
\end{equation}
we have presented the fully explicit formulae:
\begin{equation}
    \dot z = H_0 \left\{ (1+z) - \sqrt{\sum_{i=1}^N \Omega_{0i} \;(1+z)^{3(1+w_i)} } \right\};
\end{equation}
\begin{equation}
\dot y = H_0 (1-y)\left\{ 1 - (1-y) \sqrt{\sum_{i=1}^N \Omega_{0i}\; (1-y)^{-3(1+w_i)} }\right\};
\end{equation}
\begin{equation}
   \dot{y}_1 = H_0\left\{ 1 -e^{-y_1}\; \sqrt{\sum_{i=1}^N \Omega_{0i} \; e^{3y_1 (1+w_i)}}\right\} \;.
\end{equation}

This was applied initially for a $\Lambda$CDM model with four and two components, followed by an analysis with different dark energy models in Section \ref{S:dark_energy_models}. The models presented were $w_0$CDM, BAZS, linear model, CPL, logarithmic, and a couple of interactive cases. We have also presented a discussion about the power of redshift data in distinguishing different DE models from each other, concluding that, independently of how precise the future redshift drift data might be, different dark energy models cannot be distinguished purely from this type of data alone. On the other hand, RD data will certainly play an important role in constraining the free parameters from DE models and, when combined with other sources of data, might play a crucial role in helping us solve the challenging problem of understanding the nature of dark energy in the future.

\vspace{6pt}

\authorcontributions{
Conceptualization, FSNL, JPM, JS,  and MV; 
methodology, FSNL, JPM, JS,  and MV; software,  JS,  and MV; 
validation, FSNL, JPM, JS,  and MV; formal analysis, FSNL, JPM, JS,  and MV;  investigation, FSNL, JPM, JS,  and MV;  resources, FSNL and MV; 
data curation, FSNL, JPM, JS,  and MV; 
writing---original draft preparation, FSNL, JPM, JS,  and MV;  
writing---review and editing, FSNL, JPM, JS,  and MV; 
visualization, JS; supervision, MV; project administration, FSNL and MV; 
funding acquisition, FSNL and MV.
All authors have read and agreed to the published version of the manuscript.
}

\funding{\textls[-15]{
 FSNL and JPM acknowledge funding from the Funda\c c\~{a}o para a Ci\^encia e a Tecnologia (FCT, Portugal) research projects UIDB/04434/2020, UIDP/04434/2020, PTDC/FIS-OUT/29048/2017, CERN/FIS-PAR/0037/2019, EXPL/FIS-AST/1368/2021, and PTDC/FIS-AST/0054/2021. FSNL also thanks support from the FCT Scientific Employment Stimulus contract with reference CEECINST/00032/\linebreak  2018. JS was partially supported by the Hellenic Foundation for Research and Innovation (H.F.R.I.), under the “First Call for H.F.R.I. Research Projects to support Faculty members and Researchers and the procurement of high-cost research equipment" grant (Project Number: 789) and was partially supported by Taiwan's National Science and Technology Council. MV was supported by the Marsden Fund via a grant administered by the Royal Society of New Zealand.}
}

\dataavailability{All relevant data is included in the article.
}


\conflictsofinterest{The authors declare no conflict of interest.
 }

\begin{adjustwidth}{-\extralength}{0cm}
\printendnotes[custom]
\reftitle{References}

\PublishersNote{}
\end{adjustwidth}

\begin{thebibliography}{999}  
\newcommand{\arXiv}[1]{arXiv:\href{https://arxiv.org/abs/#1}{\color{blue}#1}}




\bibitem{Sandage:1962}
Sandage, A.
The change of redshift and apparent luminosity of galaxies due to the deceleration of selected expanding universes.
\emph{Astrophys. J.} \textbf{1962}, {\em 136},  319.
\url{https://doi.org/10.1086/147385}.

\bibitem{McVittie:1962}
McVittie, G.C. Appendix to the change of redshift and apparent luminosity of galaxies due to the deceleration of selected expanding universes. \emph{Astrophys. J.} \textbf{1962}, {\em136}, 334.


\bibitem{Loeb:1998}
  Loeb, A.
  Direct measurement of cosmological parameters from the cosmic deceleration of extragalactic objects.
  \emph{Astrophys. J.}  {\bf 1998}, \emph{499}, {L111.} 
\url{https://doi.org/10.1086/311375}.


  \bibitem{Linder:2008}
  Linder, E.V.
  Mapping the Cosmological Expansion.
  \emph{Rept. Prog. Phys.}  {\bf 2008}, \emph{71}, 56901.
  \url{https://doi.org/10.1088/0034-4885/71/5/056901}.


  
  \bibitem{Quercellini:2010}
  Quercellini, C.; Amendola, L.; Balbi, A.; Cabella, P.; Quartin, M.
  Real-time Cosmology.
  \emph{Phys. Rept.}  {\bf 2012}, \emph{521}, 95.
  \url{https://doi.org/10.1016/j.physrep.2012.09.002}.
 

  

  
   \bibitem{Alves:2019}
  Alves, C.S.; Leite, A.C.O.; Martins, C.J.A.P.; Matos, J.G.B.; Silva, T.A.
  Forecasts of redshift drift constraints on cosmological parameters.
  \emph{Mon. Not. Roy. Astron. Soc.}  {\bf 2019}, \emph{488}, 3607.
  \url{https://doi.org/10.1093/mnras/stz1934}.
 
  

  
  \bibitem{Liske:2008}
  Liske, J.; Grazian, A.; Vanzella, E.; Dessauges, M.; Viel, M.; Pasquini, L.; Haehnelt, M.; Cristiani, S.; Pepe, F.; \mbox{Avila, G.; et al.}
  Cosmic dynamics in the era of Extremely Large Telescopes.
  \emph{Mon. Not. Roy. Astron. Soc.}  {\bf 2008}, \emph{386}, 1192.
  \url{https://doi.org/10.1111/j.1365-2966.2008.13090.x}.
 
  \bibitem{Steinmetz:2008}
  Steinmetz, T.; Wilken, T.; Araujo-Hauck, C.; Holzwarth, R.; Hansch, T.W.; Pasquini, L.; Manescau, A.; D'odorico, S.; Murphy, M.T.; Kentischer, T.; et al.
  Laser frequency combs for astronomical observations.
  \emph{Science} {\bf 2008}, \emph{321}, 1335.
  \url{https://doi.org/10.1126/science.1161030}.

  
   
  \bibitem{Kim:2014}
 Kim, A.G.; Linder, E.V.; Edelstein, J.; Erskine, D.
  Giving Cosmic Redshift Drift a Whirl.
  \emph{Astropart. Phys.}  {\bf 2015}, \emph{62}, 195.
  \url{https://doi.org/10.1016/j.astropartphys.2014.09.004}.

  

 \bibitem{Killedar:2009}
  Killedar, M.; Lewis, G.F.
  Lyman alpha absorbers in motion: Consequences of gravitational lensing for the cosmological redshift drift experiment.
  \emph{Mon. Not. Roy. Astron. Soc.}  {\bf 2010}, 402, 650.
  \url{https://doi.org/10.1111/j.1365-2966.2009.15913.x}.

  
 \bibitem{Lazkoz:2017} Lazkoz, R.; Leanizbarrutia, I.; Salzano, V.
Forecast and analysis of the cosmological redshift drift.
\emph{Eur. Phys. J. C} \textbf{2018}, \emph{78}, 11.  
\url{https://doi.org/10.1140/epjc/s10052-017-5479-0}.

  
  \bibitem{Marcori:2018}
  Marcori, O.H.; Pitrou, C.; Uzan, J.P.; Pereira, T.S.
  Direction and redshift drifts for general observers and their applications in cosmology.
  \emph{Phys. Rev. D} {\bf 2018}, \emph{98}, 023517.
  \url{https://doi.org/10.1103/PhysRevD.98.023517}.

 

\bibitem{Lu:2022}
Lu, C.Z.; Zhang, T.; Zhang, T.J.
Statistical distribution of HI 21cm absorbers as potential cosmic acceleration probes.
\emph{{Mon. Not. R. Astron. Soc.}} {\textbf{2023}, \emph{521}, 3150--3161.}

\bibitem{Dong:2022}
Dong, C.; Gonzalez, A.; Eikenberry, S.; Jeram, S.; Likamonsavad, M.; Liske, J.; Stelter, D.; Townsend, A.;
Forecasting cosmic acceleration measurements using the Lyman-\ensuremath{\alpha} forest.
\emph{Mon. Not. Roy. Astron. Soc.} \textbf{2022}, \emph{514}, 5493--5505.
\url{https://doi.org/10.1093/mnras/stac1702}.




\bibitem{Covone:2022}
Covone, G.; Sereno, M.
Lensing cosmic drift.
\emph{Mon. Not. Roy. Astron. Soc.} \textbf{2022}, \emph{513}, 5198--5203.
\url{https://doi.org/10.1093/mnras/stac1261}.


\bibitem{Chakrabarti:2022}
Chakrabarti, S.; Gonzalez, A.H.; Eikenberry, S.; Erskine, D.; Ishak, M.; Kim, A.; Linder, E.; Nomerotski, A.; Pierce, M.; \mbox{Slosar, A.;  {et al.}}
Real-time Cosmology with High Precision Spectroscopy and Astrometry. 
\emph{{arXiv}} \textbf{2023}, {\arXiv{2203.05924}}.


\bibitem{Lu:2021}
Lu, C.Z.; Jiao, K.; Zhang, T.; Zhang, T.J.; Zhu, M.
Toward a direct measurement of the cosmic acceleration: The first preparation with FAST.
\emph{Phys. Dark Univ.} \textbf{2022}, \emph{37}, 101088.
\url{https://doi.org/10.1016/j.dark.2022.101088}.


\bibitem{Martins:2021}
Martins, C.J.A.P.; Alves, C.S.; Esteves, J.; Lapel, A.; Pereira, B.G.
Closing the cosmological loop with the redshift drift. In {Proceedings of the  The Sixteenth Marcel Grossmann Meeting on Recent Developments in Theoretical and Experimental General Relativity, Astrophysics and Relativistic Field Theories.
Online, 5--10 July 2021.}


\bibitem{Eikenberry:2019}
 Eikenberry, S.S.; Gonzalez, A.; Darling, J.; Liske, J.; Slepian, Z.; Mueller, G.; Conklin, J.; Fulda, P.; de Oliveira, C.M.; \mbox{Bentz, M.; {et al.}}
  Astro2020 Project White Paper: The Cosmic Accelerometer. 
  {\emph{arXiv} \textbf{2019}, \arXiv{1907.08271}.}
 

\bibitem{Esteves:2021}
Esteves, J.; Martins, C.J.A.P.; Pereira, B.G.; Alves, C.S.
Cosmological impact of redshift drift measurements.
\emph{Mon. Not. Roy. Astron. Soc.} \textbf{2021}, \emph{508}, L53--L57.
\url{https://doi.org/10.1093/mnrasl/slab102}.


\bibitem{Uzan:2008}
  Uzan, J.P.; Clarkson, C.; Ellis, G.F.R.
  Time drift of cosmological redshifts as a test of the Copernican principle.
 \emph{ Phys. Rev. Lett.}  {\bf 2008}, \emph{100}, 191303.
  \url{https://doi.org/10.1103/PhysRevLett.100.191303}.
 
 \bibitem{Koksbang:2022} 
Koksbang, S.M.; Heinesen, A.
Redshift drift in a universe with structure: Lema\^\i{}tre-Tolman-Bondi structures with arbitrary angle of entry of light
Phys. Rev. D \textbf{106} (2022) no.4, 043501
\url{https://doi.org/10.1103/PhysRevD.106.043501}.

 
  \bibitem{Koksbang:2015}
  Koksbang, S.M.; Hannestad, S.
  Redshift drift in an inhomogeneous universe: Averaging and the backreaction conjecture.
 \emph{J. Cosmol. Astropart. Phys. } {\bf 2016}, \emph{1601}, 9.
  \url{https://doi.org/10.1088/1475-7516/2016/01/009}.
 

\bibitem{Quartin:2009xr}
Quartin, M.; Amendola, L.
Distinguishing Between Void Models and Dark Energy with Cosmic Parallax and Redshift Drift.
\emph{Phys. Rev. D} \textbf{2010}, \emph{81}, 043522.
\url{https://doi.org/10.1103/PhysRevD.81.043522}.


\bibitem{Yoo:2010hi}
{Yoo,}  C.M.; Kai, T.; Nakao, K.i.
Redshift Drift in LTB Void Universes.
\emph{Phys. Rev. D} \textbf{2011}, \emph{83}, 43527.
\url{https://doi.org/10.1103/PhysRevD.83.043527}.


\bibitem{Geng:2015ara}
Geng, J.J.; Li, Y.H.; Zhang, J.F.; Zhang, X.
Redshift drift exploration for interacting dark energy.
\emph{Eur. Phys. J. C} \textbf{2015}, \emph{75}, 356.
\url{https://doi.org/10.1140/epjc/s10052-015-3581-8}.


\bibitem{Calabrese:2013lga}
Calabrese, E.; Martinelli, M.; Pandolfi, S.; Cardone, V.F.; Martins, C.J.A.P.; Spiro, S.; Vielzeuf, P.E.
Dark Energy coupling with electromagnetism as seen from future low-medium redshift probes.
\emph{Phys. Rev. D} \textbf{2014}, \emph{89}, 83509.
\url{https://doi.org/10.1103/PhysRevD.89.083509}.



\bibitem{Mishra:2012vi}
Mishra, P.; Celerier, M.N.; Singh, T.P.
Redshift drift as a test for discriminating between different cosmological models.
\emph{Phys. Rev. D} \textbf{2012}, \emph{86}, 83520.
\url{https://doi.org/10.1103/PhysRevD.86.083520}.



\bibitem{Li:2021}
Li, Z.; Liao, K.; Wu, P.; Yu, H.; Zhu, Z.-H.
Probing modified gravity theories with the Sandage-Loeb test.
\emph{Phys. Rev. D} \textbf{2021}, \emph{88}, 23003.
\url{https://doi.org/10.1103/PhysRevD.88.023003}.


\bibitem{Loeb:2022}
Scherrer, R.J.; Loeb, A.
Ultra Long-Term Cosmology and Astrophysics. \emph{New Astron.} \textbf{2023}, \emph{99}, 101940.

  
\bibitem{Lobo:2020}
Lobo, F.S.; Mimoso, J.P.; Visser, M.
Cosmographic analysis of redshift drift. 
\emph{J. Cosmol. Astropart. Phys. } \textbf{2020}, \emph{4}, 43.
\url{https://doi.org/10.1088/1475-7516/2020/04/043}.

  
 \bibitem{Dunsby:2015}
  Dunsby, P.K.S.; Luongo, O.
  On the theory and applications of modern cosmography.
  \emph{Int. J. Geom. Meth. Mod. Phys. }{\bf 2016}, \emph{13}, 1630002. \url{https://doi.org/10.1142/S0219887816300026}.





   \bibitem{Visser:2003}
  Visser, M.
  Jerk and the cosmological equation of state.
  \emph{Class. Quant. Grav.}  {\bf 2004}, \emph{21}, 2603.
  \url{https://doi.org/10.1088/0264-9381/21/11/006}.



\bibitem{Gibbons:2008}
{Dunajski}, M.; Gibbons, G.
Cosmic Jerk, Snap and Beyond.
\emph{Class. Quant. Grav.} \textbf{2008}, \emph{25}, 235012.
\url{https://doi.org/10.1088/0264-9381/25/23/235012}.


\bibitem{Visser:2004}
 Visser, M.
  Cosmography: Cosmology without the Einstein equations.
  \emph{Gen. Rel. Grav.}  {\bf 2005}, \emph{37}, 1541.
  \url{https://doi.org/10.1007/s10714-005-0134-8}.
  
    \bibitem{Aviles:2012}
    Aviles, A.; Gruber, C.; Luongo, O.; Quevedo, H.
  Cosmography and constraints on the equation of state of the Universe in various parametrizations.
  \emph{Phys. Rev. D} \textbf{2012}, \emph{86}, 123516. \url{https://doi.org/10.1103/PhysRevD.86.123516}.


  \bibitem{Cattoen:2007b}
  Catt\"oen, C.; Visser, M.
  The Hubble series: Convergence properties and redshift variables.
  \emph{Class. Quant. Grav.}  {\bf 2007}, \emph{24}, 5985.
  \url{https://doi.org/10.1088/0264-9381/24/23/018}.
  
  
 \bibitem{Cattoen:2008}
  Catt\"oen, C.; Visser, M.
  Cosmographic Hubble fits to the supernova data.
  \emph{Phys. Rev. D} {\bf 2008}, \emph{78,} 063501.
  \url{https://doi.org/10.1103/PhysRevD.78.063501}.
 
  
  \bibitem{jerk3}
 Eager, D.; Pendrill, A.M.; Reistad, N.
Beyond velocity and acceleration: Jerk, snap and higher derivatives.
\emph{Eur. J. Phys.} {\bf2016}, \emph{37}, 065008.
\url{https://doi.org/10.1088/0143-0807/37/6/065008}.
  
    \bibitem{Busti:2015}
  Busti, V.C.; Cruz-Dombriz, Á.; Dunsby, P.K.S.; Sáez-Gómez, D.
  Is cosmography a useful tool for testing cosmology?
  \emph{Phys. Rev. D} {\bf 2015}, \emph{92}, 123512. \url{https://doi.org/10.1103/PhysRevD.92.123512}.
  
  \bibitem{Lusso:2020}
    Risaliti, G.; Lusso, E.
  Cosmological constraints from the Hubble
diagram of quasars at high redshifts.
   \emph{Nat. Astr.} \textbf{2019}, \emph{3}, 272--277. \url{https://doi.org/10.1038/s41550-018-0657-z}.
  
  \bibitem{Hu:2022}
  Hu, J.P.; Wang, F.Y.
  High-redshift cosmography: Application and comparison with different methods.
   \emph{Astron. Astrophys.} \textbf{2022}, \emph{661}, A71. \url{https://doi.org/10.1051/0004-6361/202142162}.
   
   \bibitem{Yang:2020}
    Yang, T.; Banerjee, A.; Colgáin, E.Ó.
  Cosmography and flat $\Lambda$CDM tensions at high redshift.
   \emph{Phys. Rev. D} \textbf{2020}, \emph{102}, 123532. \url{https://doi.org/10.1103/PhysRevD.102.123532}.
   
   \bibitem{Capozziello:2020}
    Capozziello, S.; D’Agostino, R.; Luongo, O.
  High-redshift cosmography: Auxiliary variables versus Padé polynomials.
   \emph{Mon. Not. R. Astron. Soc.} \textbf{2020}, 494,  2576–2590. \url{https://doi.org/10.1093/mnras/staa871}.
   
 
  
 



\bibitem{Madsen:1992}
Madsen, M.S.; Ellis, G.F.R.; Mimoso, J.P.; Butcher, J.A.
Evolution of the density parameter in inflationary cosmology reexamined.
\emph{Phys. Rev. D} \textbf{1992}, \emph{46}, 1399--1415.
 \url{https://doi.org/10.1103/PhysRevD.46.1399}


\bibitem{PDG:2022} 
Particle Data Group. 
Available online: \url{https://pdg.lbl.gov/2020/reviews/rpp2020-rev-astrophysical-constants.pdf} { (accessed on 21 March. 2024).} 

\bibitem{Workman:2022}
Particle Data Group; Workman, R.L.; Burkert, V.D.; Crede, V.; Klempt, E.; Thoma, U.; Tiator, L.; Agashe, K.; Aielli, G.; \mbox{Allanach, B.C.; et al.}
Review of Particle Physics.
\emph{Prog. Theor. Exp. Phys.} \textbf{2022}, \emph{2022}, 083C01.
\url{https://doi.org/10.1093/ptep/ptac097}.


\bibitem{Planck2018}
Aghanim, N.; Akrami, Y.; Ashdown, M.; Aumont, J.; Baccigalupi, C.; Ballardini, M.; B.; ay, A.J.; Barreiro, R.B.; Bartolo, N.; \mbox{Basak, S.; et al.} Planck 2018 results. VI. Cosmological parameters.  \emph{Astron. Astrophys.} \textbf{2020}, \emph{641}, A6.

\bibitem{Pantheon+}
Brout, D.; Scolnic, D.; Popovic, B.; Riess, A.G.; Carr, A.; Zuntz, J.; Kessler, R.; Davis, T.M.; Hinton, S.; Jones, D.; et al. The Pantheon+ Analysis: Cosmological Constraints.  
\emph{{Astrophys.~J.}} {\textbf{2022}, \emph{938}, 110.}

\bibitem{last-scattering}
Recombination (cosmology), 
 {Available} online: \url{https://en.wikipedia.org/wiki/Recombination_(cosmology)} 
 {(accessed on 21 March 2024).}


\bibitem{Corless:1996}
  Corless, R.; Gonnet, G.; Hare, D.; Jeffrey, D.; Knuth, D.
  On the Lambert W function.
  \emph{Adv. Comput. Math.} \textbf{1996}, \emph{5},  329--359. \url{https://doi.org/10.1007/BF02124750}.
  
\bibitem{Valluri:2000}
  Valluri, S.R.; Jeffrey, D.J.; Corless, R.M.
  Some applications of the Lambert W function to physics.
  \emph{Can. J. Phys.}  {\bf 2000}, \emph{78}, 823.
  \url{https://doi.org/10.1139/p00-065}.
 
  
\bibitem{Valluri:2009}
  Valluri, S.R.; Gil, M.; Jeffrey, D.J.; Basu, S.
  The Lambert W function and quantum statistics.
  \emph{J. Math. Phys.}  {\bf 2009}, \emph{50}, 102103.
  \url{https://doi.org/10.1063/1.3230482}.
  



 
\bibitem{Boonserm:2013}
 Boonserm, P.; Ngampitipan, T.; Visser, M.
  Regge-Wheeler equation, linear stability, and greybody factors for dirty black holes.
  \emph{Phys. Rev. D} {\bf 2013}, \emph{88}, 041502.
  \url{https://doi.org/10.1103/PhysRevD.88.041502}.
  
  

  
  \bibitem{Sonoda:2013a}
  Sonoda, H.; 
  Solving renormalization group equations with the Lambert $W$ function.
  \emph{Phys. Rev. D} {\bf 2013}, \emph{87}, 85023.
  \url{https://doi.org/10.1103/PhysRevD.87.085023}.



\bibitem{Visser:2018-LW}
  Visser, M.
  Primes and the Lambert W function.
  \emph{Mathematics} {\bf 2018}, \emph{6}, 56. 
  \url{https://doi.org/10.3390/math6040056} .
  
  


\bibitem{Charters:2009ku}
Charters, T.; Mimoso, J.
Self-interacting scalar field cosmologies: Unified exact solutions and symmetries.
\emph{J. Cosmol. Astropart. Phys.}  \textbf{2010}, \emph{8}, 22.
\url{https://doi.org/10.1088/1475-7516/2010/08/022}.

\bibitem{Balbi:2007fx}
  Balbi, A.; Quercellini, C.
  The time evolution of cosmological redshift as a test of dark energy,
   \emph{ Mon. Not. Roy. Astron. Soc.} \textbf{2007}, \emph{382}, 1623. 
  \url{https://doi.org/10.1111/j.1365-2966.2007.12407.x} 

  \bibitem{Corasaniti:2007}
  Corasaniti, P.S.; Huterer, D.; Melchiorri, A.
  Exploring the dark energy redshift desert with the Sandage-Loeb test.
    \emph{Phys. Rev. D}  \textbf{2007}, 75, 062001.
  \url{https://doi.org/10.1103/PhysRevD.75.062001}. 

  \bibitem{BAZS}
   Barboza, E.M., Jr.;  Alcaniz, J.S.; ; Zhu, ; Z.-H.; ; Silva, R.
  Generalized equation of state for dark energy.
    \emph{Phys. Rev. D} \textbf{2009}, 80, 043521.
  \url{https://doi.org/10.1103/PhysRevD.80.043521}.


\bibitem{CP}
  Chevallier, M.; Polarski, D.
  Accelerating Universes with Scaling Dark Matter.
    \emph{Int. J. Mod. Phys. D} \textbf{2001}, \emph{10}, 213.
  \url{https://doi.org/10.1142/S0218271801000822}. 

\bibitem{L}
  Linder, E.V.
  Exploring the Expansion History of the Universe.
    \emph{Phys. Rev. Lett.} \textbf{2003}, \emph{90}, 091301.
  \url{https://doi.org/10.1103/PhysRevLett.90.091301}.

\bibitem{Efs}
 Efstathiou, G.
  Constraining the equation of state of the Universe from distant Type Ia supernovae and cosmic microwave background anisotropies.
    \emph{Mon. Not. Roy. Astron. Soc.} \textbf{1999}, \emph{310},  842–850.
  \url{https://doi.org/10.1046/j.1365-8711.1999.02997.x}.


\bibitem{int1}
 Cao, S.; Liang, N.
  Interaction between dark energy and dark matter: Observational constraints from OHD, BAO, CMB and SNe Ia.
   \emph{ Int. J. Mod. Phys. D} \textbf{2013},  \emph{22}, 1350082.
  \url{https://doi.org/10.1142/S021827181350082X}.

\bibitem{int2}
Xia, D.-M.; Wang, S.
Constraining interacting dark energy models with latest cosmological observations.
\emph{Mont. Not. Roy. Astron. Soc.} \textbf{2016},  463, 952–956.
\url{https://doi.org/10.1093/mnras/stw2073}.


\bibitem{DET5}
Zhang, J.F.; Zhang, M.; Jin, S.J.; Qi, J.Z.; Zhang, X.
Cosmological parameter estimation with future gravitational wave standard siren observation from the Einstein Telescope.
\emph{J. Cosmol. Astropart. Phys.}  \textbf{2019}, \emph{9}, 68.
\url{https://doi.org/10.1088/1475-7516/2019/09/068}.

\bibitem{DET4}
Qi, J.Z.; Jin, S.J.; Fan, X.L.; Zhang, J.F.; Zhang, X.
Using a multi-messenger and multi-wavelength observational strategy to probe the nature of dark energy through direct measurements of cosmic expansion history.
\emph{J. Cosmol. Astropart. Phys.}  \textbf{2019}, 12, 42.
\url{https://doi.org/10.1088/1475-7516/2021/12/042}.

\bibitem{DET3}
Zhang, M.; Wang, B.; Wu, Pe.; Qi, Ji.; Xu, Y.; Zhang, Ji.; Zhang, X.
Prospects for Constraining Interacting Dark Energy Models with 21 cm Intensity Mapping Experiments.
\emph{Astrophys. J.} \textbf{2021}, \emph{918},  56.
\url{https://doi.org/10.3847/1538-4357/ac0ef5}.

\bibitem{DET2}
Jin, S.J.; Zhu, R.Q.; Wang, L.F.; Li, H.L.; Zhang, J.F.; Zhang, X.
Impacts of gravitational-wave standard siren observations from Einstein Telescope and Cosmic Explorer on weighing neutrinos in interacting dark energy models.
\emph{Commun. Theor. Phys.} \textbf{2022}, \emph{74}, 105404.
\url{https://doi.org/10.1088/1572-9494/ac7b76}.


\bibitem{DET1}
Hou, W.T.; Qi, J.Z.; Han, T.; Zhang, J.F.; Cao, S.; Zhang, X.
Prospects for constraining interacting dark energy models from gravitational wave and gamma ray burst joint observation.
\emph{J. Cosmol. Astropart. Phys.}  \textbf{2023}, \emph{5}, 17.
\url{https://doi.org/10.1088/1475-7516/2023/05/017}.


\bibitem{codex}
CODEX Phase A, Science Case, document E-TRE-IOA-573-0001 Issue 1 (2010).\\
The science case for CODEX,
an ultra-stable high-resolution spectrograph for the E-ELT.\\
Available online at \url{http://research.iac.es/proyecto/codex//media/codex_files/CO01_Science_Case.pdf} (accessed on 21 March 2024).

  

\end{thebibliography}
\end{document}